\documentclass[fleqn,usenatbib]{mnras}

\usepackage{xcolor}
\usepackage{arydshln}
\usepackage{pdflscape}
\usepackage{lscape}
\usepackage{afterpage}
\usepackage{booktabs}
\usepackage{braket}
\usepackage{mathptmx,amsmath}
\usepackage{txfonts}

\usepackage[T1]{fontenc}
\usepackage{ae,aecompl}

\usepackage{graphicx}	% Including figure files
\usepackage{tablefootnote,longtable,threeparttable}
\newcommand{\mearth}{M_\oplus}
\newcommand{\rearth}{R_{\rm \oplus}}

\def\ms{\hbox{\,m\,s$^{-1}$}}         %m.s -1
\def\m2s2{\hbox{\,m$^{2}$\,s$^{-2}$}} %m2.s -2
\title{Biases in retrieving planetary signals
in the presence of quasi-periodic stellar activity }

\author[M. Damasso et al.]{
M. Damasso,$^{1}$\thanks{E-mail: mario.damasso@inaf.it}
M. Pinamonti,$^{1}$
G. Scandariato$^{2}$
and A. Sozzetti$^{1}$
\\
$^{1}$INAF - Osservatorio Astrofisico di Torino, Via Osservatorio 20, I-10025 Pino Torinese, Italy\\
$^{2}$INAF - Osservatorio Astrofisico di Catania, Via S. Sofia 78, I-95123 Catania
}

\date{Accepted 2019 August 3. Received 2019 August 2; in original form 2019 March 19}

\pubyear{2018}

\begin{document}
\label{firstpage}
\pagerange{\pageref{firstpage}--\pageref{lastpage}}
\maketitle

\begin{abstract}
Gaussian process regression is a widespread tool used to mitigate stellar correlated noise in radial velocity time series. It is particularly useful to search for and determine the properties of signals induced by small-size, low-mass planets ($R<4\rearth$, $m<10\mearth$). By using extensive simulations based on a quasi-periodic representation of the stellar activity component, we investigate the ability in retrieving the planetary parameters in 16 different realistic scenarios. We analyse systems composed by one planet and host stars having different levels of activity, focusing on the challenging case represented by low-mass planets, with Doppler semi-amplitudes in the range 1-3 $\ms$). We consider many different configurations for the quasi-periodic stellar activity component, as well as different combinations of the observing epochs. We use commonly-employed analysis tools to search for and characterize the planetary signals in the datasets. The goal of our injection-recovery statistical analysis is twofold. First, we focus on the problem of planet mass determination. Then, we analyse in a statistical way periodograms obtained with three different algorithms, in order to explore some of their general properties, as the completeness and reliability in retrieving the injected planetary and stellar activity signals with low false alarm probabilities. This work is intended to provide some understanding of the biases introduced in the planet parameters inferred from the analysis of radial velocity time series that contain correlated signals due to stellar activity. It also aims to motivate the use and encourage the improvement of extensive simulations for planning spectroscopic follow-up observations.
\end{abstract}

\begin{keywords}
planetary systems -- methods: numerical -- methods: statistical -- stars: activity -- techniques: radial velocities
\end{keywords}

%%%%%%%%%%%%%%%%%%%%%%%%%%%%%%%%%%%%%%%%%%%%%%%%%%

%%%%%%%%%%%%%%%%% BODY OF PAPER %%%%%%%%%%%%%%%%%%

\section{Introduction}
Over the last two decades, ground and space-based transit surveys have unveiled the very rich and unexpected variety of distant worlds orbiting stars of different spectral type, age and activity levels in the Milky Way Galaxy. Moreover, starting or upcoming surveys will dramatically increase the number of planet detections and the variety of the statistical planetary sample within the next decade. Most of the currently known exoplanets have been discovered through transit observations, which alone provide only the size of the planets. Unless transit timing variations (TTV) are detected (e.g. \citealt{agol17}), the planetary masses have to be measured through the analysis of the radial velocity (RV) time variations of the host star, induced by the gravitational interaction with the planets. 

For transiting extrasolar planets with measured masses, the bulk densities can be determined, and through the analysis of the planetary mass-radius diagram one can infer the bulk composition and structure of observed planets, and compare them with formation and evolution models. This in turn helps to put reliable constrains on the different migration mechanisms that have been proposed to describe the observed architectures of the planetary systems. Moreover, precise measurements of the bulk density can help in the prediction of the surface temperature of potentially habitable planets.     

One of the most interesting results obtained thanks to the discoveries made by the Kepler/K2 mission is that the radii of small, close-in extrasolar planets (R < 4 $\rearth$ and P < 100 days) follow a bi-modal distribution \citep{owenwu13,fulton17,zeng17a,zeng17b}. Different studies have confirmed and strengthened this finding by improving the precision on the host star radii \citep{fulton18,berger18}. A gap located at $\sim$1.5-2$\rearth$ divides a population of super-Earth/Earth-size planets from that of sub-Neptunes. This evidence could be the direct proof that photoevaporation is a major mechanism responsible for planetary mass loss, as predicted by some theoretical studies (e.g. \citealt{owenwu13,lopez14}). The smallest planets below the gap could have lost the outer gaseous envelope through a X-UV radiation-driven erosion, while those with larger radii could still retain a significant fraction of volatiles. In this context, the smallest planets are expected to have the highest bulk densities, indicative of a rocky composition, while planets above the gap should have lower densities as a consequence of their H$_{\rm 2}$/He gaseous envelopes (gas dwarfs) or H$_{\rm 2}$O-dominated ices/fluids (water worlds), according to two of the mainstream scenarios (e.g. \citealt{owen18,wu18,zeng19}).

The discovery of the 'radius gap' for small-size planets makes the accurate and precise measurement of their masses a still more urgent matter.
Measuring the masses of small-size, low-mass planets through the radial velocity method is not an easy task for several reasons. First of all, currently only few high-resolution and high stability spectrographs in use for several years (e.g. HARPS, HARPS-N, HIRES) are able to provide the RV precision necessary for precise measurements of the RV amplitude of low-mass exoplanets. %An extensive use of such instruments by individual teams is not reliably possible, preventing the possibility to observe planet host stars for a long time span and with frequent visits.
The 'weighing' of low-mass planets is made an even more complex task by crucial aspects, such as the observing conditions, the magnitude and spectral type of the host star, the orbital period of the planet, and the level of the stellar activity, which can have a major negative impact on the analyses (e.g. \citealt{dumusque16,fisher16,dumusque17}). All these issues must be carefully accounted for when planning observing campaigns aimed at characterizing extrasolar planets, and some preparatory, target-tailored analyses are usually performed to define the best observing strategy to measure masses with the best precision possible \citep[e.g.][]{damasso18}.

In this study we define a framework that could help the selection of the most promising targets for a spectroscopic follow-up with chances to provide exoplanet masses with good accuracy and precision. We investigate reliable scenarios typically faced by teams involved in characterization studies of exoplanets, and test tools for data analysis that are presently in the toolbox of every exoplanet hunter.  
We use extensive simulations of RV datasets, described in Sect. \ref{sect:simsetup}, for two types of analyses. In the first part of the paper (Sect. \ref{sec:mcanalysis}) we use statistics to investigate how accurately the planetary signals are retrieved in presence of a quasi-periodic stellar activity component. Here we consider: \textit{i}) systems with only one planet; \textit{ii}) different levels of stellar activity; \textit{iii}) stellar rotation periods $P_{\rm rot}$ close to or well-separated from the planetary orbital period $P_{\rm orb}$; \textit{iv}) short and long evolutionary time scales of the correlated stellar activity signal included in the RV time series; \textit{v}) the case of a survey carried out with only one spectrograph during one and two observing seasons, in order to simulate a scenario representative of on-going and future planet characterization campaigns such as those focused on Kepler/K2 or TESS targets (the inclusion of RV measurements collected with more than one instrument would introduce a higher level of complexity that we decided to avoid in this work, nevertheless without making our simulations less realistic). The choice of simulating planets with $P_{\rm orb}\sim P_{\rm rot}$ originates from the intrinsic difficulties encountered in separating the stellar activity from the planetary signal in real RV datasets, when they have very similar periods \citep[e.g.][]{vander16}.

The blind detection of small-amplitude planetary signals is another challenging aspect related to RV time series affected by stellar activity. Simulated RV time series, containing signals due to stellar activity, have been used in previous works to test techniques for detecting planets (e.g. \citealt{feng16,dumusque17}), or to apply a variety of methods to estimate the Bayesian evidence for a range of planetary signals \citep{nelson18}. \cite{pinamontietal2017} studied the efficiency of different period-search approaches in retrieving small-amplitude planetary signals, making use of simulated datasets also containing simple modeling of the stellar activity. In this perspective, we exploit the mock datasets for a complementary analysis aimed at characterizing features appearing in the periodograms that can be related to the different properties of the simulated quasi-periodic stellar activity scenarios. This second part of our study is described in Sect. \ref{sect:periodanalysis}. As done by \cite{pinamontietal2017}, the frequency content of each dataset is investigated by means of three common and publicly available softwares: the Generalized Lomb-Scargle periodograms (GLS, \citealt{zech09}), the Bayesian Generalised Lomb-Scargle (BGLS, \citealt{mortieretal2015}) and the FREquency DEComposer (FREDEC, \citealt{baluev2013}). The results of this analysis can be useful to interpret the morphology of the periodograms calculated on real data, which usually are the first to be investigated when searching for planetary signals.
%--------------------------------------------------------------------------------------

\section{Simulation set-up}% (1): generation of the mock datasets}
\label{sect:simsetup}

%\textit{General context.}

In this work we aim to explore a number of realistic scenarios through a statistical analysis of a large number of samples.
We simulated systems consisting of only one planet to keep the size of the parameter space limited, since the inclusion of additional planets would require a much larger set of simulations in order to cover many system architectures and draw statistically meaningful conclusions. We assume that all the data have been collected with the same instrument. In what follows, we summarize the main features that we have taken into account to simulate the RV datasets, and provide the justification for their adoption.

\subsection{Planetary Doppler signals}
We simulated planets on circular orbits, which is a reasonable assumption for low-mass planets with low $P_{\rm orb}$ \citep{vaneylen19}. The semi-amplitude of the Doppler planetary signal $K_{\rm p}$ was kept fixed to 1 $\ms$ for all the cases. The planetary orbital periods $P_{\rm orb}$ were generated between $\sim$10 and 20 days, either close or far apart from the stellar rotation period $P_{\rm rot}$ (see Sect. \ref{subsec:stactmodel}).   
Assuming $K_{\rm p}$=1 \ms and K-to-mid M dwarfs as host stars\footnote{In this work we do not assume stars of some specific spectral type, since this information is not crucial for the simulation set-up. In general, we consider a typical star observed by space-based transit surveys as a representative target, i.e. a main sequence star of spectral type between G and M, sufficiently bright to be selected for high-precision spectroscopic follow-up.}, the corresponding planetary masses are in the range $1.5 - 4.0$ $M_\oplus$. Using the relations of \cite{weiss14} to provide likely estimates for the planet sizes, this mass interval corresponds to radii in the range 1.1-1.5 $R_\oplus$. This is the interval that includes the first mode of the bi-modal distribution of close-in planet sizes \citep{fulton17,fulton18}.  
Therefore, we simulated systems that are particularly challenging for the state-of-the-art high-precision and high-stability spectrographs that can be accessed nowadays. Nonetheless, they are among the more interesting scenarios for exoplanet characterization studies. 

\subsection{Stellar activity model}
\label{subsec:stactmodel}
The stellar component in the RVs was simulated as a quasi-periodic signal, which is described by the kernel with the same name used in Gaussian process (GP) regression analysis. The quasi-periodic GP model is often used to effectively correct (mitigate) the stellar activity term in RV time series that contain signals modulated on $P_{\rm rot}$ \citep[e.g.][]{affer16,damasso17,pinamontietal2018}, and the effective use of GP models in recovering small-amplitude planetary signals was demonstrated for the case of synthetic RV time series \citep[e.g.][]{dumusque17}. As often happens, the stellar rotation period can be measured from space- or ground-based photometry, or from time series of spectroscopic activity indicators. A quasi-periodic model can be assumed as a realistic, even if not necessarily complete, representation of the stellar activity contribution due to active regions. Our simulations do not have the same level of detail as those devised by \cite{dumusque16}, which also include contributions such as those due to stellar oscillations, granulation, and supergranulation. Nonetheless, by injecting a quasi-periodic signal we are simulating a reliable scenario useful for a statistical study, despite being only one of several possible representations of the stellar component in RV time series.

The quasi-periodic kernel used in our analyses is described by the following covariance function and hyper-parameters:
\begin{equation}
\label{eq_gpker}
\mathcal{K}(t, t') = h^2\times \exp \bigg[-\frac{(t-t')^2}{2{\tau_{\rm AR}}^2} - \frac{\sin^{2}(\dfrac{\pi(t-t')}{P_{\rm rot}})}{2w^2}\bigg] + \sigma^{2}_{\rm RV}(t)\times\delta_{\rm t, t'},
\end{equation}
where $\mathcal{K}$(t, t$^{\prime}$) represents the covariance matrix element between observation at time \textit{t} and \textit{t$^{\prime}$}, $h$ is the amplitude of the correlations; $P_{\rm rot}$ is the stellar rotation period of the star; $w$ is the length scale of the periodic component, describing the level of high-frequency variation within a complete stellar rotation; $\tau_{\rm AR}$ is the correlation decay time scale, that can be physically related to the lifetime of the active regions (AR); and $\sigma_{\rm RV}$(t) is the RV uncertainty at epoch $t$ (see next Section).

We explored two different ranges of $P_{\rm rot}$, by adopting $P_{\rm rot}\sim$10 days for active stars and $P_{\rm rot}\sim$20 days for the low-activity sample. This choice is rather arbitrary but based on the known fact that more active stars tend to rotate faster \citep[e.g.][]{stelzer16,mascareno18}.

Among the GP hyper-parameters, $\tau_{\rm AR}$ is particularly tricky, as it is not clear whether in general it carries some real physical information. This is true especially when a consistency check is not possible using contemporary photometric observations. For instance, several examples exist of RV datasets we have analysed using a quasi-periodic GP model for which estimates of $\tau_{\rm AR}$ have values close to $P_{\rm rot}$ \citep[e.g.][]{damasso18,pinamontietal2018,haywood18,malavolta18}. On the other hand, other studies cast doubts on this hypothesis, stating that much longer time scales solutions should be more physically valid, especially for M dwarfs \citep[e.g.][and references therein]{perger17}. We explore here two different regimes for $\tau_{\rm AR}$, by simulating rapidly evolving correlations, $\tau_{\rm AR}\sim$ $P_{\rm rot}$, and stars with activity evolving over longer time scales, $\tau_{\rm AR}\sim$ 10$\times P_{\rm rot}$.

As for the amplitude of the stellar signals, $h$, we investigated two representative regimes of stellar activity. We defined them on the basis of the typical scatter observed in the RVs\footnote{Referring to the measurements collected with the kind of spectrographs considered in this work.} for a low-activity star, $h \sim$3 \ms, and a high-activity star, $h\sim$15 \ms (e.g. \citealt{santos2000,mascareno17,mascareno18,damasso19}). 
In a more general sense, our definition of low and high stellar activity levels should be intended as relative to the small semi-amplitude of the planetary signal investigated in this work, i.e. $h \gtrsim K_{\rm p}$ and $h \gg K_{\rm p}$.
%Our simulations represent what should be expected from the stellar activity of typical transit exoplanets hosts.% Analysis of RV data collected simultaneously with space-based photometry have been attempted only recently \citep{oshagh17}, and there is not yet a sufficient amount of RV data for different targets to derive empirical relations between the photometric and spectroscopic scatters. 

\subsection{RV uncertainties}
We assumed that the RV measurements are collected by a high-resolution and high-stability spectrograph, as HARPS or HARPS-N, in order to simulate a realistic follow-up campaign in terms of access to the observing facilities.
We took into account the expected average precision for such instruments\footnote{This can be estimated through the on-line exposure time calculators available at \url{https://www.eso.org/observing/etc/bin/gen/form?INS.NAME=HARPS+INS.MODE=spectro} and \url{http://www.tng.iac.es/instruments/harps/}} and for a typical target eligible for follow-up observations (see e.g. \citealt{fisher16}). For instance, the RV precision expected with HARPS-N for a K0V star of magnitude $V$=10.5, with 1800 seconds of exposure taken at airmass 1.2 and with a seeing of 1.2 arcsec, is $\sigma_{\rm RV}\sim$ 1 \ms.

However, to simulate a more realistic final error budget we adopted a lower average data precision of 2 \ms, which is equivalent to considering an additional noise term $\sim$1.7 \ms added in quadrature to the nominal uncertainty. This can realistically represent what is expected for fainter targets, and takes into account possible worsening factors such as the cross-correlation function (CCF) noise, the rotational broadening of the CCF, and additional noise dependent on the specific RV-extraction algorithm. We generated the uncertainty for each data point from a normal distribution with mean equal to 2 \ms and standard deviation equal to 0.3 \ms. 

\subsection{Observing seasons and epochs}
%As a consequence of the different astrophysical scenarios described so far, a total of eight possible combinations are explored. This number is doubled since 
We simulated up to two consecutive observing seasons (or semesters). We analysed the time series with one and two seasons of observations as two separate cases, investigating the significance of the retrieved activity and planetary parameters as a function of the number of measurements.

To make the simulations realistic, the epochs of observation for the first semester have been generated adopting a typical schedule of nights with HARPS-N at the TNG telescope in La Palma allocated to the GTO collaboration\footnote{https://plone.unige.ch/HARPS-N/science-with-harps-n} and to the GAPS/GAPS2 programmes \citep{benatti16}, which are organized in shared risk mode. In our case, we adopted the real allocated observing time between the end of October 2017 and the end of March 2018, consisting in 63 nights. %Hereafter we describe how the arrays of epochs (or time stamps) used to simulate groups of 100 different mock RV datasets have been generated, starting with the case of one observing season.
We assumed that the target was visited once per night, and for each night we generated randomly an observing time from a normal distribution with mean equal to the midnight of that date and standard deviation equal to 0.01 days. Then, taking into account any loss due to bad weather, or Moon contamination constraints, or technical problems, we assumed a duty cycle equal to $65\%$\footnote{This duty cycle is the typical percentage of useful time for the GAPS programme after five years of observations.} and randomly drawn 40 out of 63 epochs, which we define hereafter as the $N_{\rm epochs, s1}$ ensemble. 

In the case of two semesters of observations, we added 40 additional time stamps to $N_{\rm epochs,\:s1}$. These new time stamps were randomly drawn from an array of equally spaced epochs delayed by 365 days with respect of the first semester. As before, the epochs were randomly shifted around the midnight of the date. Then, each final dataset of epochs used to simulate two semesters consists of $N_{\rm epochs,\:s2}$=80. \textbf{Fig. \ref{fig:simdatasets} shows examples of randomly drawn $N_{\rm epochs, s1}$ and $N_{\rm epochs, s2}$ time stamps.} 

This procedure was repeated 50 times, resulting in 50 groups of simulated RV datasets, both for one and two observing seasons. Each group is composed of 100 RV time series that share the same time stamps, \textbf{with 5\,000 simulated datasets for each scenario, and 80\,000 mock datasets in total}. By varying the epochs within each group, coupled to the randomly drawn values of $P_{\rm orb}$ and $T_{\rm 0,b}$, we made the results independent from a particular choice of the observing sampling, as discussed in \ref{appx2}, despite the epochs are drawn from the same 'mold' calendar. A more detailed investigation of the effects due to different samplings is beyond the scope of this paper.   

\subsection{Global model and generation of the mock datasets}
\begin{figure}
   \centering
   \includegraphics[width=9cm]{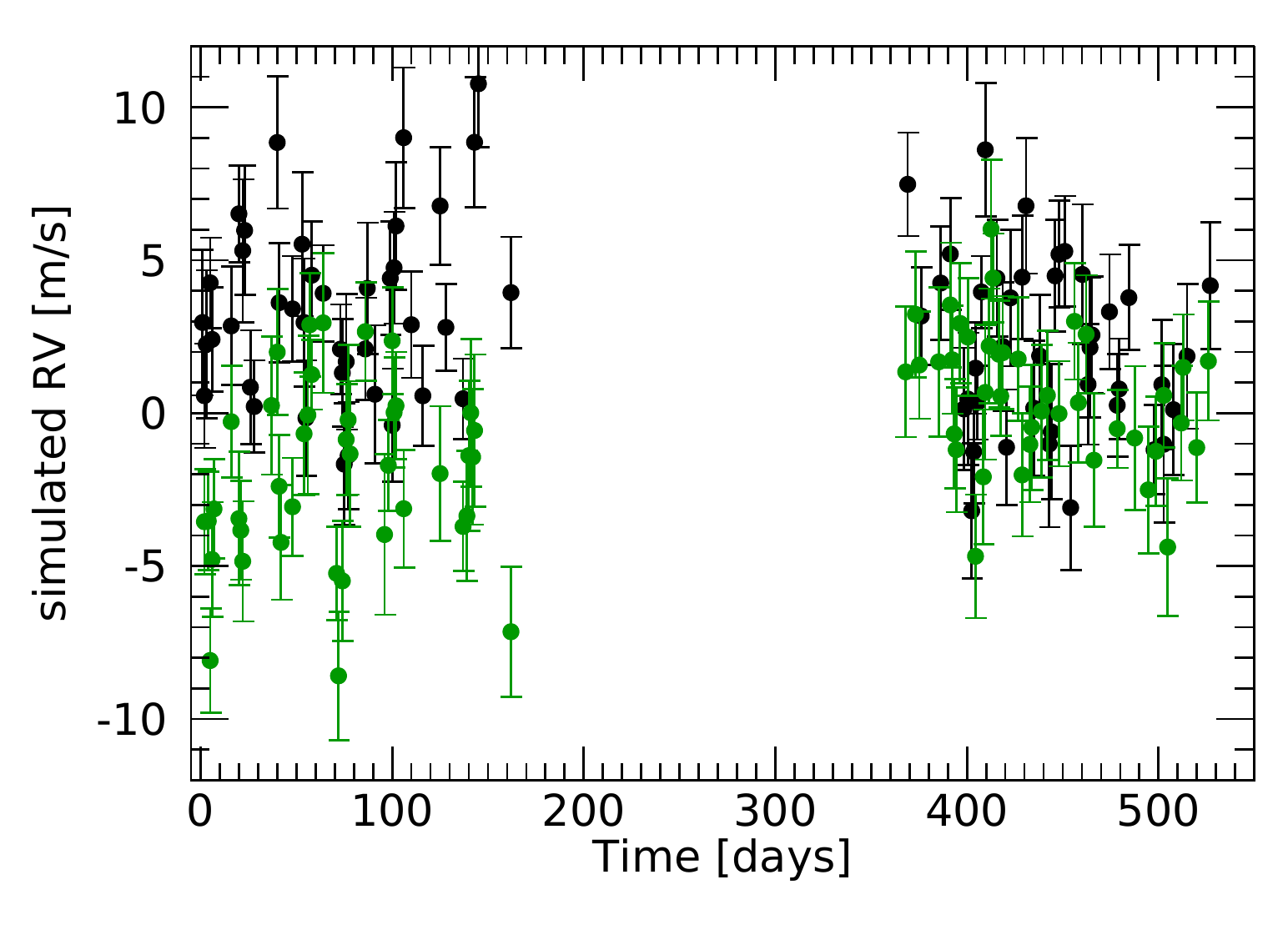}
   \includegraphics[width=9cm]{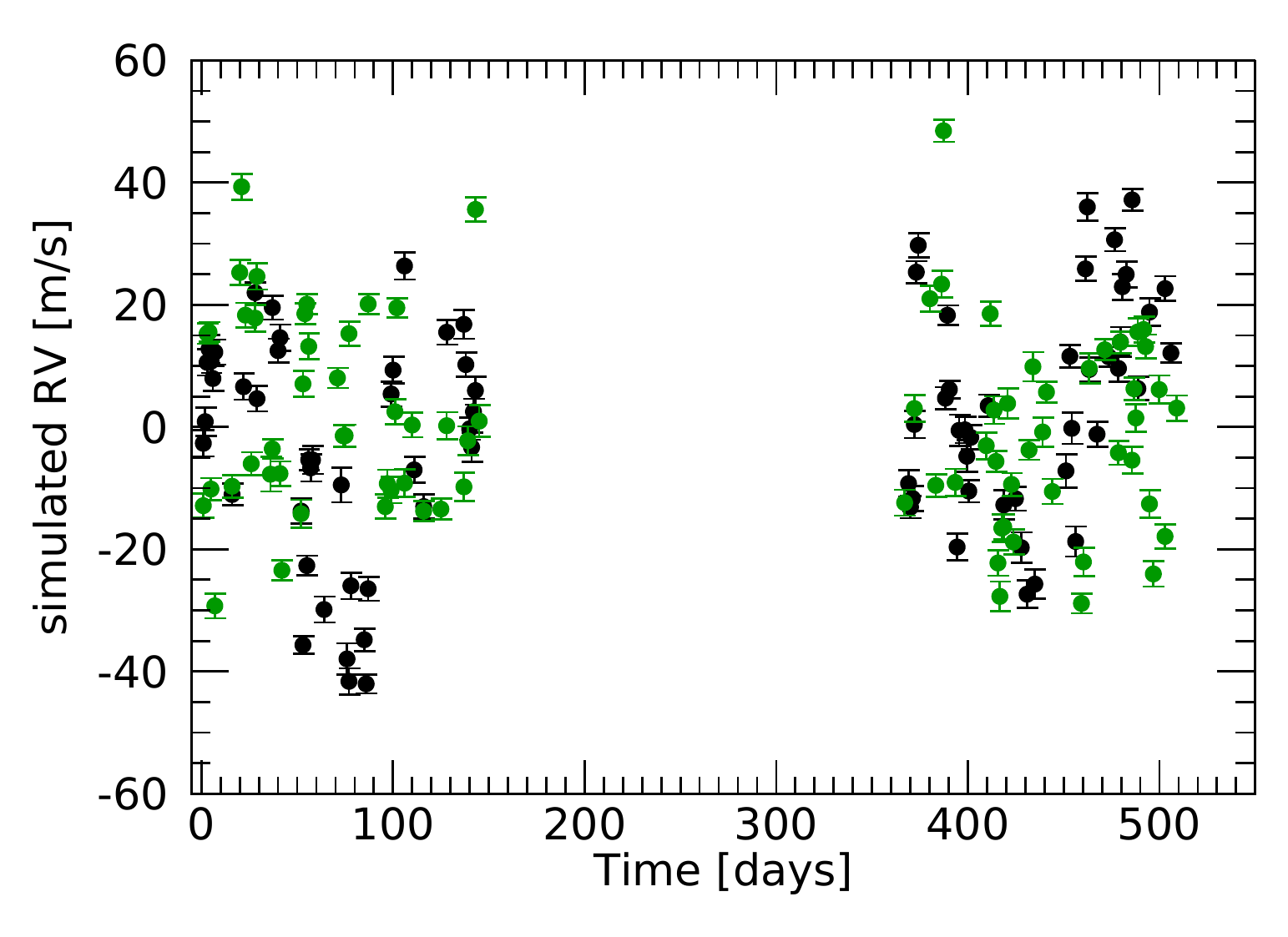}
      \caption{Examples of mock RV datasets used in this work.\textit{Upper panel:} Two seasons of data for low-activity stars with $P_{\rm rot}\ne P_{\rm orb}$ and $\tau_{\rm AR} \gg P_{\rm rot}$. \textit{Lower panel:} Two seasons of data for active stars with $P_{\rm rot}\sim P_{\rm orb}$ and $\tau_{\rm AR}\sim P_{\rm rot}$.}
         \label{fig:simdatasets}
\end{figure}

The final model used to simulate each dataset is the sum of one circular orbit (characterized by semi-amplitude $K_{\rm b}$, period $P_{\rm b}$, and time of inferior conjunction $T_{\rm 0,b}$) and a correlated quasi-periodic signal representing the stellar activity contribution. For each simulated time series, the model parameters have been randomly drawn from distributions as shown in Table \ref{Table:simsetup}, depending on the specific astrophysical scenario. Then, after injecting the two signals, each data point of the time series was randomly shifted within its error bar $\sigma_{\rm RV}$(t) (Table \ref{Table:simsetup}), using a normal distribution centred on zero and with $\sigma=\sigma_{\rm RV}$(t). Examples of simulated datasets are shown in Fig. \ref{fig:simdatasets}. 
It is worth noticing that stellar and planetary parameters used to simulate the one season datasets and the corresponding two seasons datasets are not the same, since we were not interested in studying the effect of additional observations on a star-by-star basis, but only in a statistical sense.

\begin{table}
  \caption[]{Simulation set-up: how each parameter has been drawn to generate the total 80\,000 datasets simulated in this work.}
         \label{Table:simsetup}
         \centering
         \normalsize
   \begin{tabular}{cccc}
            \hline
            \hline
            \noalign{\smallskip}
            \textbf{Parameter}   & \textbf{Low activity} & \textbf{High activity} \\
            \noalign{\smallskip}
            \hline
            %\noalign{\smallskip}
            %Stellar activity GP model & &\\
            \noalign{\smallskip}
            $h$ [m\,$s^{-1}$] & $\mathcal{N}$(3,0.5$^{2}$) &  $\mathcal{N}$(15,1$^{2}$) \\ 
            \noalign{\smallskip}
            $P_{\rm rot}$ [days] & $\mathcal{N}$(20,1.5$^{2}$) & $\mathcal{N}$(10,1$^{2}$) \\ 
            \noalign{\smallskip}
            $\tau_{\rm AR}$ [days] & \multicolumn{2}{c}{$\mathcal{N}$($\mu_{\rm P_{\rm rot}}$,$\sigma_{\rm P_{\rm rot}}^{2}$) [short $\tau_{\rm AR}$]} \\ 
            & \multicolumn{2}{c}{$\mathcal{N}$(10$\times\mu_{\rm P_{\rm rot}}$,10$^{2}$) [long $\tau_{\rm AR}$]} \\ 
            \noalign{\smallskip}
            $w$ & \multicolumn{2}{c}{$\mathcal{N}$(0.5,0.2$^{2}$)}  \\ 
            \noalign{\smallskip}
            \hline
            %\noalign{\smallskip}
          %  \noalign{\smallskip}
         %   $\sigma_{\rm jit}$ [m\,$s^{-1}$] & $\mathcal{N}$(2,0.5$^{2}$) & $\mathcal{N}$(3,0.5$^{2}$) \\ 
          %  \noalign{\smallskip}
         %   \hline
         %   \noalign{\smallskip}
            %Planetary orbital parameters & &\\
            \noalign{\smallskip}
            $K_{\rm b}$ [m\,s$^{-1}$] & \multicolumn{2}{c}{1, fixed} \\ 
            \noalign{\smallskip}
            $P_{\rm b}$ [days] & \multicolumn{2}{c}{$\mathcal{N}$($\mu_{\rm P_{\rm rot}}$,$\sigma_{\rm P_{\rm rot}}^{2}$) [$P_{\rm orb}\sim P_{\rm rot}$]} \\ 
            \noalign{\smallskip}
            & $\mathcal{N}$(13,1.5$^{2}$) & $\mathcal{N}$(18,1.5$^{2}$) \\
            & [$P_{\rm orb}$ $\neq$ $P_{\rm rot}$] & [$P_{\rm orb}$ $\neq$ $P_{\rm rot}$] \\
            \noalign{\smallskip}
            $T_{\rm 0,b}$ [days] & \multicolumn{2}{c}{$\mathcal{N}$(50,5$^{2}$)} \\ 
            \noalign{\smallskip}
            \hline
            %\noalign{\smallskip}
            %Internal RV uncertainties & &\\
            \noalign{\smallskip}
            $\sigma_{\rm RV}$ [m\,s$^{-1}$] & \multicolumn{2}{c}{$\mathcal{N}$(2,0.3$^{2}$)} \\ 
            \noalign{\smallskip}
            \hline
            \hline
     \end{tabular}    
\end{table}

A key point in our simulations is that, in each mock dataset, the stellar activity term was randomly drawn using the \texttt{sample} function of the \texttt{GP} object implemented in the \texttt{GEORGEv0.2.1} package \citep{george14} used to define the GP framework. Once the GP hyper-parameters have been drawn, the \texttt{sample} function returns a randomly drawn list of predictions at the time stamps of the observations\footnote{Details can be found in http://betatim.github.io/posts/gaussian-processes-with-george/}. This allows for simulations free from a particular representation of the stellar activity component given a set of hyper-parameters and, within our working framework, this procedure makes our results statistically representative of what it is expected on average for an activity signal described by a quasi-periodic model.
%We note that the stellar and planetary parameters used to simulate the $N_{\rm epochs, s2}$ RV time series are not the same of the corresponding N$_{\rm epochs, s1}$ datasets. This implies that our results describe what it is expected \textit{on average} for each specific scenario, and not on a star-by-star basis, i.e. what it would be by using the same activity and planetary parameters for one and two semesters.
\\
 
%Table \ref{Table:simuldistr} summarizes the properties related to the posterior distributions of combinations of the input parameters.
The general properties of the distributions of the simulated parameters are summarized in Table \ref{Table:simuldistr}. To define the final set of mock RV time series to be analysed, we adopted selection criteria in order to avoid overlaps (interchange) between samples with $P_{\rm rot}\sim P_{\rm orb}$ and those with $P_{\rm rot}\neq P_{\rm orb}$, so that they represent two distinct groups. We imposed constraints on |$P_{\rm rot}$-$P_{\rm orb}$|: concerning samples with $P_{\rm rot}\sim P_{\rm orb}$ we considered only those for which |$P_{\rm rot}$-$P_{\rm orb}$|<4 days, while for samples with $P_{\rm rot}\neq P_{\rm orb}$ we analysed only samples for which |$P_{\rm rot}$-$P_{\rm orb}$|>4 days. This selection resulted in a very low percentage of rejected samples for the case of active stars (<1.5$\%$), while between 6$\%$ and 8$\%$ of the mock datasets were removed from the analysis for stars with low level of activity, ensuring a statistically large number of RV time series.
  
% \textbf{NOTA. Alcuni esempi di segnali stellari in sistemi reali analizzati (che non siano stelle super-attive, come TAP26 o V830 Tau): M69, semi-ampiezza fittata $\pm$35 \ms; GJ504, $\pm$30 \ms; K2-36, $\pm$ 15 \ms. }

\section{Monte Carlo fitting analysis}
\label{sec:mcanalysis}
We first analysed the simulated datasets described in Sect. \ref{sect:simsetup} from the perspective of someone interested in determining the mass of a transiting planet, within the framework of GP regression to model the stellar activity signal.
Hereafter we describe the set-up used for a Monte Carlo (MC) analysis of the simulated RV datasets, then we summarize the results schematically by distinguishing among different scenarios.

\subsection{Analysis set-up}
\label{sect:simsetup2} 
Conservatively, we assumed that the planet orbital period is known from the transit analysis with a precision of 0.5 days, and that $P_{\rm rot}$ can be constrained with a precision of 5 days from photometry and/or from other activity diagnostics, as those derived from high-resolution spectra. We fit the planetary signal with a Keplerian with null eccentricity, and the stellar signal using the same GP kernel described in Eq. \ref{eq_gpker}. Our working hypothesis is that the choice of a quasi-periodic kernel is justified, so that this representation can be used confidently to model the correlated stellar activity signal present in the RV time series. We supposed that the amplitude $h$ of the stellar activity term cannot be well constrained a-priori, but it could be as high as 30 $\ms$ for an active star. For the evolutionary time scale $\tau_{\rm AR}$ we adopted a large uninformative prior, since it is usually difficult to constrain this hyper-parameter a priori (an alternative prior choice is discussed in Sect. \ref{subsec:alternprior}). 
All these assumptions are at the basis of our choice of the prior distributions used in the MC analysis (Table \ref{Table:simpriors}). %Moreover, $\tau_{\rm AR}$ can also strongly depend from the actual timespan of the RV measurements, so that any prior information available from ancillary data collected in previous epochs could not be sufficiently informative.

\begin{table}
  \caption[]{Priors used to recover the model parameters through a Monte Carlo sampling combined with a Gaussian process regression.}
         \label{Table:simpriors}
         \centering
         %\normalsize
   \begin{tabular}{cccc}
            \hline
            \hline
            \noalign{\smallskip}
            \textbf{Jump parameter}   & \textbf{Low activity} & \textbf{High Activity} \\
            \noalign{\smallskip}
            \hline
            %\noalign{\smallskip}
            %Stellar activity hyper-parameters & &\\
            \noalign{\smallskip}
            $h$ [m\,s$^{-1}$] & $\mathcal{U}$(0, 10) &  $\mathcal{U}$(0, 30) \\ 
            \noalign{\smallskip}
            $\tau_{\rm AR}$ [days] & \multicolumn{2}{c}{$\mathcal{U}$(0, 1000)} \\ 
            \noalign{\smallskip}
            $w$ &  \multicolumn{2}{c}{$\mathcal{U}$(0, 1)} \\ 
            \noalign{\smallskip}
            $P_{\rm rot}$ [days] & \multicolumn{2}{c}{$\mathcal{U}$($P_{\rm rot,\:sim.}$-5, $P_{\rm rot,\:sim.}$+5)} \\ 
            \noalign{\smallskip}
            \hline
            %\noalign{\smallskip}
            %Planetary orbital parameters & &\\
            \noalign{\smallskip}
            $K_{\rm b}$ [m\,s$^{-1}$] & \multicolumn{2}{c}{$\mathcal{U}$(0, 5)} \\ 
            \noalign{\smallskip}
            $P_{\rm orb}$ [days] & \multicolumn{2}{c}{$\mathcal{U}$(P$_{\rm orb, simul.}$-0.5, P$_{\rm orb, simul.}$+0.5)} \\ 
            \noalign{\smallskip}
            $T_{\rm 0, b}$ [days] &  \multicolumn{2}{c}{$\mathcal{U}$(0, 100)} \\ 
            \noalign{\smallskip}
            \hline
            \noalign{\smallskip}
            $\sigma_{\rm jit}$ [\ms] &  \multicolumn{2}{c}{$\mathcal{U}$(0, 10)} \\
            \noalign{\smallskip}
            \hline
            \hline
     \end{tabular}    
\end{table}

Our fitting framework includes a constant jitter term $\sigma_{\rm jit}$ which is added in quadrature to the uncertainty $\sigma_{\rm RV}(t)$, and takes into account any additional source of uncorrelated noise, instrumental and astrophysical. Doing this we followed a common practice, therefore we included $\sigma_{\rm jit}$ in our analysis even if our simulated data were built without considering any additional uncorrelated noise, that is $\sigma_{\rm jit}$=0 $\ms$ in the mock RVs by construction. It is an interesting exercise to explore whether the fitting procedure results in non-negligible estimates for $\sigma_{\rm jit}$ and for which scenario this potentially occurs.

The data analysis was carried out with commonly-used and well-tested tools to perform Monte Carlo sampling within a Gaussian process framework. We used the open source Bayesian inference tool \texttt{MultiNestv3.10} (e.g. \citealt{feroz13}), through the \texttt{pyMultiNest} \texttt{python} wrapper \citep{buchner14}, with 500 live points and a sampling efficiency of 0.8, \textbf{which is a recommended value for the purpose of parameter estimation}. The GP module is the same publicly available \texttt{GEORGEv0.2.1} \texttt{python} library we used to generate the datasets.  

%---------------------------------------------------------------------------------------

\subsection{Model parameter retrieval}
\label{sect:results}
We examined the outcomes of the MC analyses by studying separately the results for each simulated scenario. To asses their quality, we used the quantities defined in Table \ref{Table:analysedparam}. We adopt the 16$^{th}$, 50$^{th}$, and 84$^{th}$ percentiles of their posterior distributions as our statistics.

\begin{table}
  \caption[]{Description of quantities analysed in this work and derived from the Monte Carlo sampling combined with a Gaussian process regression (Sect. \ref{sect:results} and Table \ref{Table:mcmcanalysis}).}
         \label{Table:analysedparam}
         \centering
         %\small
   \begin{tabular}{lp{0.35\textwidth}}
            \hline
            \hline
            \noalign{\smallskip}
            \textbf{Quantity}  & \textbf{Description} \\
            \noalign{\smallskip}
            $h_{\rm ratio}$ & Ratio between the retrieved amplitude of the activity term and the corresponding injected value for each mock RV dataset;\\
            \noalign{\smallskip}
            $P_{\rm rot,\: ratio}$ & Ratio between the retrieved stellar rotation period in the activity term and the corresponding injected value for each mock RV dataset;\\
            \noalign{\smallskip}
            $\tau_{\rm AR, ratio}$ & Ratio between the retrieved active region evolutionary time scale and the corresponding injected value for each mock RV dataset;\\
            \noalign{\smallskip}
            $\sigma_{\rm ratio}$ & Average of the ratio between the total error budget in the retrieved model and the injected uncertainty (at epoch $t$) for each mock RV dataset (see Eq. \ref{eq_sigma_rat});\\
            \noalign{\smallskip}
            $K_{\rm b,\:ratio\: 50\%}$ & Ratio between the 50$^{\rm th}$ percentile of the planetary semi-amplitude retrieved from each mock dataset (i.e. the best-fit value obtained for $K_{\rm b}$) and the corresponding injected semi-amplitude;\\
            \noalign{\smallskip}
            $K_{\rm b,\:ratio\: 68.3\%}$ & Ratio between the 68.3$^{\rm th}$ percentile of the planetary semi-amplitude $K_{\rm b}$ retrieved from each mock dataset and the corresponding injected value;\\
            \noalign{\smallskip}
             $K_{\rm b,\:50\%}$/$\sigma^{\rm -}_{\rm K_{\rm b}}$ & Ratio between the 50$^{\rm th}$ percentile of the posterior of the planetary semi-amplitude $K_{\rm b}$ and its lower uncertainty. We assume this parameter to quantify the significance of the retrieved planetary Doppler semi-amplitude with respect to a null value.\\
            \noalign{\smallskip}
            \hline
            \hline
     \end{tabular}    
\end{table}

% the ratios between the best-fit and injected values for the hyper-parameters $h$, $\theta$ and $\tau_{\rm AR}$ of the GP quasi-periodic kernel; K$_{\rm b,ratio, 50\%}$=$\frac{K_{\rm b, retrieved}}{K_{\rm b, injected}}$, defined as the ratio, for each dataset, between the 50$^{th}$ percentile of the distribution for the retrieved semi-amplitude of the planetary signal, and the corresponding injected semi-amplitude K$_{\rm b, injected}$=1 \ms. This ratio defines the \textit{accuracy} of the derived planetary signal;  K$_{\rm b}$/$\sigma^{\rm -}_{\rm K_{\rm b}}$, defined as the ratio between the values K$_{\rm b, retrieved}$ and the corresponding lower uncertainty. This quantifies the significance of the retrieved planetary Doppler semi-amplitude with respect to a null value.  
The results are summarized in Table \ref{Table:mcmcanalysis}. We remind that, for each simulated scenario, the quantiles are calculated over all the selected mock RV time series, and they represent the average expectations over 50 different realizations of the observing epochs.
Hereafter we discuss separately the different scenarios, as well as highlight selected interesting outcomes of our analysis.

\begin{figure}
   \centering
   \includegraphics[width=8cm,height=6.5cm]{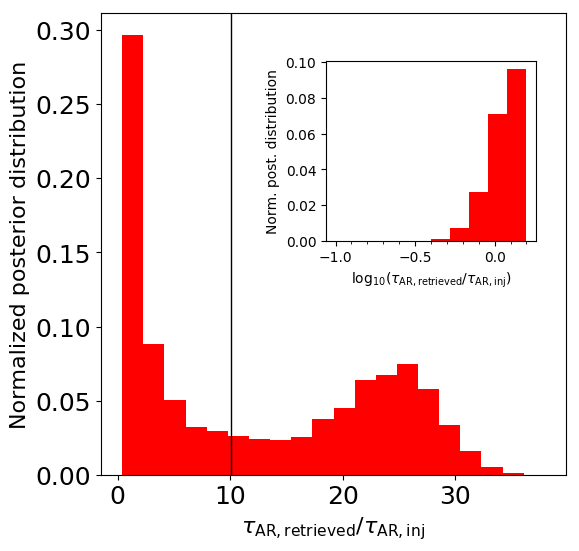}
   \includegraphics[width=8cm,height=6.5cm]{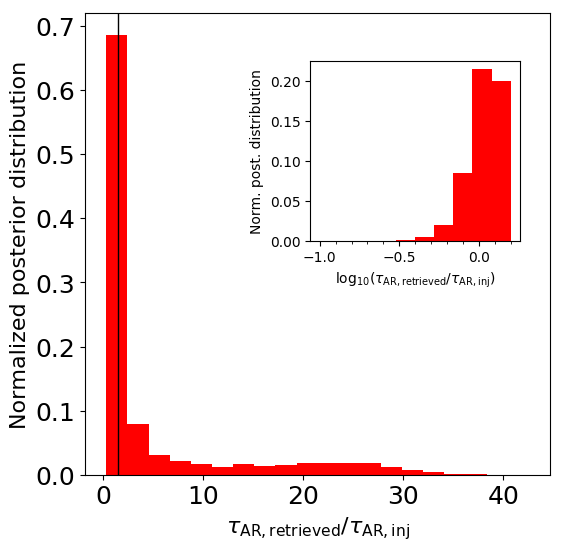}
      \caption{Posterior distributions for $\tau_{\rm AR,ratio}$=$\frac{\tau_{\rm AR, retrieved}}{\tau_{\rm AR, injected}}$ for the scenario represented by a star with low and quickly variable activity (short $\tau_{\rm AR}$) hosting a planet with $P_{\rm orb}\sim P_{\rm rot}$. The upper and lower panels show the results for one and two seasons of data, respectively. The vertical lines indicate the median of each distribution. \textbf{The inset plots are a zoomed in view around $\tau_{\rm AR,ratio}$=1 in log-scale, showing that there are very few cases where $\tau_{\rm AR}$ is underestimated.} Similar results were obtained for the case $P_{\rm orb}\neq P_{\rm rot}$}
         \label{fig:taucomp1}
\end{figure}

\subparagraph{Case I. Low-activity star; $P_{\rm orb}\sim P_{\rm rot}$.} Concerning the planetary term $K_{\rm b, ratio\: 50\%}$, which represents the accuracy on the retrieved value of $K_{\rm b}$, it increases by $\sim20\%$ after a second season of observations, but the detection significance $K_{\rm b,\:50\%}$/$\sigma^{\rm -}_{\rm K_{\rm b}}$ remains unchanged and stays below 2$\sigma$ (therefore implying a non detection) even with two seasons of data. There is evidence for a broadening of its distribution toward higher values for the scenario with long $\tau_{\rm AR}$ ($K_{\rm b,\:50\%}$/$\sigma^{\rm -}_{\rm K_{\rm b}}\sim3$ at +1$\sigma$), for which the length-scale of the stellar activity correlated signal is longer than $P_{\rm rot}$.
Concerning the stellar component, with only one season of data $\tau_{\rm AR}$ comes strongly overestimated when $\tau_{\rm AR}\sim P_{\rm rot}$. The accuracy improves after two seasons, even though a tail at longer $\tau_{\rm AR}$ persists, as can be seen in Fig. \ref{fig:taucomp1}. A similar tendency is seen when $\tau_{\rm AR}\gg P_{\rm rot}$, but the overestimate is smaller and the posterior distributions for $\tau_{\rm AR}$ appear more symmetrical. 
This result can be partially explained taking into account the small amplitude of the stellar component, which makes the signal reconstruction more difficult in presence of activity evolving on short-term time scales, especially with a scarce number/sampling of data (cfr. with \textit{case III}). We show in Sect. \ref{subsec:alternprior} that a different choice of the prior on $\tau_{\rm AR}$ has a positive impact on the accuracy of the retrieved values.

\begin{figure}
   \centering
   \includegraphics[width=9cm]{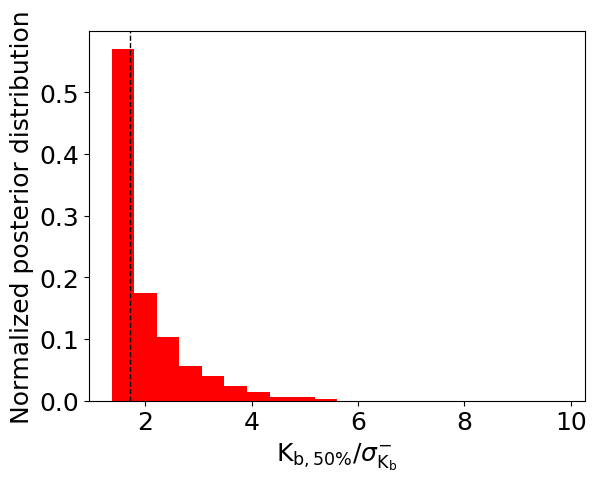}
   \includegraphics[width=9cm]{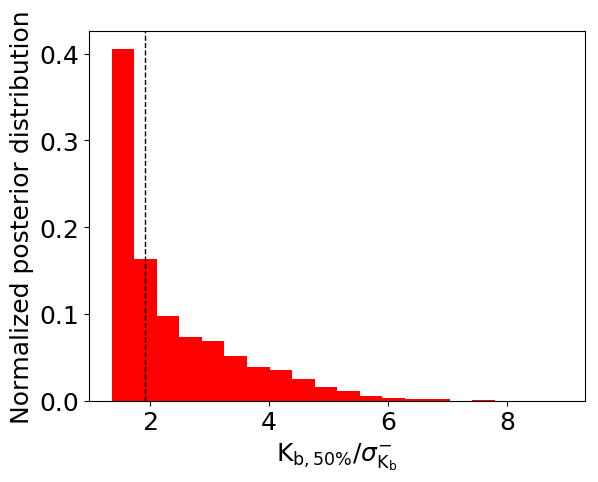}
      \caption{Posterior distributions for the planet detection significance $K_{\rm b,\:50\%}$/$\sigma^{\rm -}_{\rm K_{\rm b}}$  concerning the scenario represented by a star with low and slowly variable activity (long $\tau_{\rm AR}$) hosting a planet with $P_{\rm orb} \neq P_{\rm rot}$. The upper plot shows the result for one season of data, and the lower plot that for two seasons of observations. The vertical lines indicate the median of each distribution.}
         \label{fig:ksigma2}
   \end{figure}

\subparagraph{Case II. Low-activity star; $P_{\rm orb}\neq P_{\rm rot}$.} The main outcome is that, while the detection significance generally stays below 2$\sigma$, it increases to $\sim2\sigma$ for the case $\tau_{\rm AR}\gg$P$_{\rm rot}$ with two seasons of data, with the distribution moving toward higher values (Fig. \ref{fig:ksigma2}). This result represent a slight improvement with respect the corresponding scenario where $P_{\rm orb}\sim P_{\rm rot}$, suggesting that there should be an effect due to how close the stellar rotation and orbital periods are. The distribution of $K_{\rm b, ratio\: 50\%}$ remains unaffected passing from  $N_{\rm epochs, s1}$ to  $N_{\rm epochs, s2}$ observations, and there are no appreciable differences for the stellar component compared to \textit{Case I}, especially for the distributions of $\tau_{\rm AR}$ that are similar to those in Fig. \ref{fig:taucomp1}.

\subparagraph{Case III. Active star; $P_{\rm orb}\sim P_{\rm rot}$.} The semi-amplitude $K_{\rm b}$ is overestimated by 100$\%$ or more when $\tau_{\rm AR}\sim P_{\rm rot}$, even with $N_{\rm epochs,\:s2}$ data. The overestimate of $K_{\rm b,\:50\%}$ decreases to 20$\%$ for the case with $N_{\rm epochs,\:s2}$ data and long $\tau_{\rm AR}$, but with a relative uncertainty of $\sim66\%$. The significance $K_{\rm b,\:50\%}$/$\sigma^{\rm -}_{\rm K_{\rm b}}$, which is $\sim1.5-1.6\sigma$ in all cases, has narrower distributions for $\tau_{\rm AR}\sim P_{\rm rot}$, while for $\tau_{\rm AR}\gg P_{\rm rot}$ the distributions shift toward higher values, i.e. toward slightly more significant detections. Concerning the stellar component, the hyper-parameter $\tau_{\rm AR}$ is well recovered after two observing seasons (indicating that an uninformative prior performs well in these cases), while with one season of data and for $\tau_{\rm AR}\sim P_{\rm rot}$ the distribution of $\tau_{\rm AR}$ is skewed toward values larger than unity, but to a less extent than for quiet stars, confirming that the level of activity determines how accurately $\tau_{\rm AR}$ is fitted (as discussed for \textit{Case I}). 

\subparagraph{Case IV. Active star; $P_{\rm orb}\neq P_{\rm rot}$.} While $K_{\rm b}$ is overestimated by 100$\%$ with only one season of data and for $\tau_{\rm AR}\sim P_{\rm rot}$, it becomes more accurate for stationary stellar activity ($\tau_{\rm AR}\gg P_{\rm rot}$). The overestimate observed for short $\tau_{\rm AR}$ reduces to 50$\%$ by doubling the number of RVs. The detection significance stays below 2$\sigma$, but it begins to increase sensibly for stars with $\tau_{\rm AR}\gg P_{\rm rot}$ and $N_{\rm epochs, s2}$ observations, similarly to what is observed for low-activity stars. As for \textit{Case III}, the hyper-parameter $\tau_{\rm AR}$ is well recovered after two observing seasons.   

\subparagraph{Upper limits as defined by the 68$^{\rm th}$ percentile.} Due to the small amplitude of the planetary signal injected in the simulated datasets, the posterior distributions of $K_{\rm b}$ for each single RV dataset tend to peak typically to very low values close to 0 $\ms$. When working with real planetary systems, in such cases the 68.3$^{\rm th}$ percentile is commonly used to provide an upper limit for $K_{\rm b}$ instead of considering the median of the posterior distribution. Therefore, for each scenario we analysed the distributions $K_{\rm b,\:ratio\: 68.3\%}$ of the ratio between the 68.3$^{\rm th}$ percentile of the $K_{\rm b}$ posterior and the injected semi-amplitude, for each dataset. From Table \ref{Table:mcmcanalysis} we note that the 68.3$^{\rm th}$ percentile appears as a more accurate estimate for $K_{\rm b}$ than the 50$^{\rm th}$ percentile for low-active stars when $P_{\rm orb}\ne P_{\rm rot}$ and with $N_{\rm epochs,\:s2}$ observations. All things being equal, the same conclusion is also valid for active stars but only for the cases with long evolutionary time scales $\tau_{\rm AR}$.  

\subparagraph{The role of the uncorrelated jitter.} One issue often debated when fitting RV time series is how the results for the uncorrelated jitter $\sigma_{\rm jit}$ should be interpreted. As mentioned in Section \ref{sect:simsetup2}, we did not include any uncorrelated noise term when building the simulated RV datasets, but we use it as a free parameter $\sigma_{\rm jit}$ in the fitting analysis. 
To quantify the relevance of the uncorrelated jitter in determining our results, we define the ratio:
\begin{equation}
\label{eq_sigma_rat}
    \sigma_{\rm ratio} = \left \langle {\frac{\sqrt[]{\sigma_{\rm RV}^{\rm 2}(t)+\sigma_{\rm jit}^{2}} }{\sigma_{\rm RV}(t)}} \right \rangle
\end{equation}
for each mock dataset and consider the posterior distribution of this quantity for each simulated scenario (Table \ref{Table:mcmcanalysis}). 
For a low-activity star $\sigma_{\rm ratio}$ is very close to one, especially for the case with longer $\tau_{\rm AR}$, pointing out that the fitted $\sigma_{\rm jit}$ is negligible, in agreement with the injected $\sigma_{\rm jit}$=0. This effect is even more evident after two observing seasons. Even though the activity term is not well fitted for the cases with short $\tau_{\rm AR}$, the amplitude of the stellar signal is comparable to the average uncertainty of the RVs, implying that an additional uncorrelated noise is not required to improve the fit, which is likely limited by the sampling.   

For stars with high activity and short $\tau_{\rm AR}$, with $N_{\rm epochs,\:s1}$ data the fitted $\sigma_{\rm jit}$ determines a total error budget which is on average typically twice the uncertainty, increasing nearly up to four times $\sigma_{\rm RV}(t)$ at 1$\sigma$ level. For the sample with long $\tau_{\rm AR}$ the distribution of $\sigma_{\rm ratio}$ is much closer to one. Besides the effects of a larger amplitude of the stellar activity term, this result reveals that there is also a dependence of the fitted $\sigma_{\rm jit}$ from $\tau_{\rm AR}$. In fact, with only one season of data $\tau_{\rm AR}$ is poorly fitted, especially for the cases with short $\tau_{\rm AR}$, and this corresponds to the highest values for $\sigma_{\rm ratio}$ (with some influence due to the choice of the prior for $\tau_{\rm AR}$). The highest $\sigma_{\rm jit}$ correspond to the cases where $P_{\rm rot,\: ratio}$ is retrieved with less precision. These results point out that a higher uncorrelated jitter occurs in presence of a large stellar activity signal, in particular that higher uncorrelated jitter values can be expected for active stars with activity changing over short time scales.     

\subsubsection{Selected scenarios with injected $K_{\rm b}$>1 $\ms$}
\label{subsec:higherk}
\begin{figure}
   \centering
   \includegraphics[width=9cm]{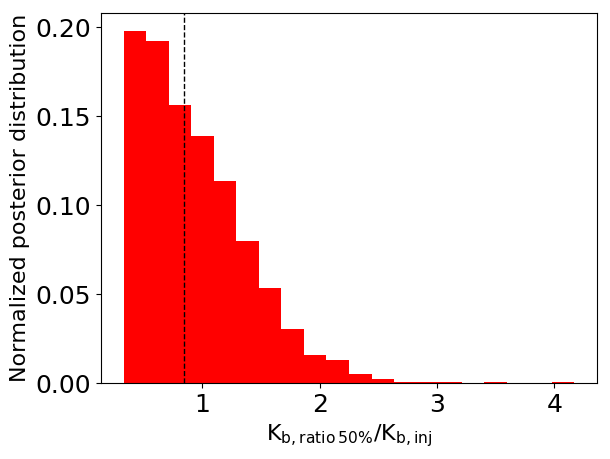}
   \centering
   \includegraphics[width=9cm]{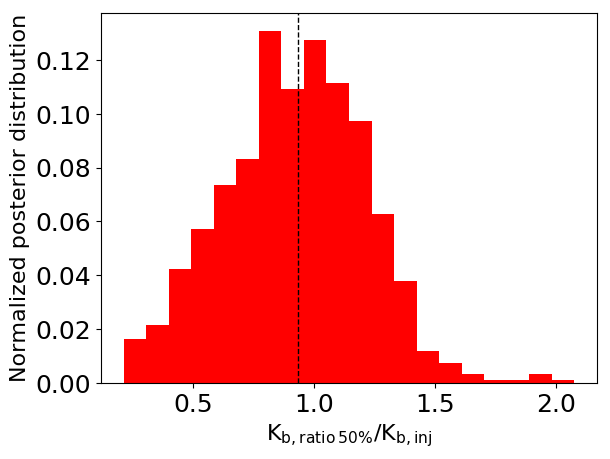}
      \caption{Distributions of the posterior $K_{\rm b,ratio\:50\%}$=$\frac{K_{\rm b, retrieved}}{K_{\rm b, injected}}$ for the scenario represented by a star with low and stable activity (long $\tau_{\rm AR}$) hosting a planet with P$_{\rm orb}\neq$P$_{\rm rot}$. The results refer to the case of datasets composed by one semester of observations. The injected planetary signals have  $K_{\rm b, inj}$=1 $\ms$ (upper plot), and $K_{\rm b, inj}$=2 $\ms$ (lower plot). The vertical lines indicate the median of each distribution.}
         \label{fig:kcomp1}
   \end{figure}

\begin{figure}
   \centering
   \includegraphics[width=9cm]{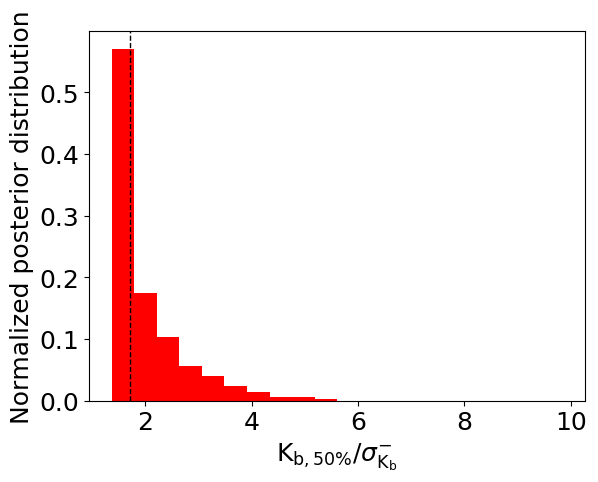}
   \centering
   \includegraphics[width=9cm]{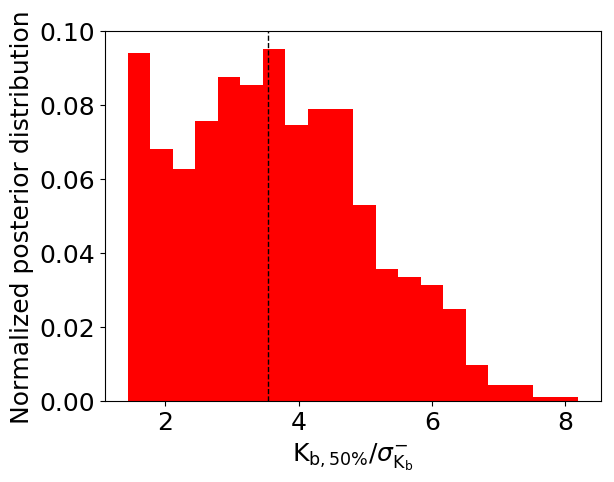}
      \caption{Posterior distributions of the quantity $K_{\rm b,\:50\%}$/$\sigma^{\rm -}_{\rm K_{\rm b}}$ for the scenario represented by a star with low and stable activity (long $\tau_{\rm AR}$) hosting a planet with $P_{\rm orb}\neq P_{\rm rot}$. The results refer to the case of datasets composed by one semester of observations. The injected planetary signals have $K_{\rm b, inj}$=1 $\ms$ (upper plot), and $K_{\rm b, inj}$=2 $\ms$ (lower plot). The vertical lines indicate the median of each distribution.}
         \label{fig:ksigcomp1}
   \end{figure}

Using a lower number of mock datasets, we conducted an exploratory analysis on some selected scenarios, limited to $N_{\rm epochs,\:s1}$ data, to see how the results change by injecting planetary signals with semi-amplitudes greater than 1 \ms. We have simulated 1\,000 mock RV datasets, corresponding to 10 groups with different time sampling realizations, each composed of 100 RV time series. We first consider the likely best-case scenario for detecting the planetary signal, represented by a star with low and stable activity (long $\tau_{\rm AR}$) hosting a planet with $P_{\rm orb}\neq P_{\rm rot}$, by injecting a Keplerian with $K_{\rm b}$=2 $\ms$. In this case we get $K_{\rm b,ratio\:50\%}$=0.9$\pm$0.3 (68.3$^{\rm th}$ percentile=1.04) and $K_{\rm b,\:50\%}$/$\sigma^{\rm -}_{\rm K_{\rm b}}$=3.5$\pm$1.5. Compared with the same scenario where $K_{\rm b, inj}$=1 $\ms$, the posterior distribution for $K_{\rm b,ratio\:50\%}$ appears nearly symmetrical around 0.9-1 $\ms$, while for $K_{\rm b, inj}$=1 \ms the distribution is asymmetric, with the peak shifted toward $\sim$0.5 $\ms$ (Fig. \ref{fig:kcomp1}). The typical significance of the detection is twice than for the case with $K_{\rm b, inj}$=1 $\ms$, with a relative uncertainty of $\sim$50$\%$ (Fig. \ref{fig:ksigcomp1}). 

For the likely worst-case scenario, represented by an active star with rapidly variable activity (short $\tau_{\rm AR}$) hosting a planet with $P_{\rm orb}\sim P_{\rm rot}$, we injected a signal with $K_{\rm b}$=3 $\ms$, and retrieved $K_{\rm b,ratio\:50\%}$=0.7$^{\rm +0.2}_{\rm -0.1}$ (68.3$^{\rm th}$ percentile=1.01) and $K_{\rm b,\:50\%}$/$\sigma^{\rm -}_{\rm K_{\rm b}}$=1.57$^{\rm +0.09}_{\rm -0.05}$. While the significance of the detection does not change with respect to the case $K_{\rm b}$=1 $\ms$, the distribution for $K_{\rm b,ratio\:50\%}$ is symmetrical with a median below 1, pointing out that, even though the semi-amplitude is on average underestimated by 30$\%$, it is closer to the true value. 

Finally, we explored the case of a quiet star with short $\tau_{\rm AR}$ hosting a planet with $P_{\rm orb}\sim P_{\rm rot}$ and $K_{\rm b}$=2 $\ms$. We get $K_{\rm b,ratio\:50\%}$=0.9$^{\rm +0.5}_{\rm -0.4}$ (68.3$^{\rm th}$ percentile=1.09) and $K_{\rm b,\:50\%}$/$\sigma^{\rm -}_{\rm K_{\rm b}}$=1.9$^{\rm +1.4}_{\rm -0.3}$. These results indicate that the planet is retrieved with higher precision and increasing significance with respect to the case where $K_{\rm b}$=1 \ms.

An interesting result, common to all the three cases analysed here, is that the 68.3$^{\rm th}$ percentile is substantially equal to the injected $K_{\rm b}$. Looking at Table \ref{Table:mcmcanalysis}, when $K_{\rm b}$=1 $\ms$ the same situation happens only for the scenarios \textit{i}) low activity stars and decoupled periodicities, with $K_{\rm b}$ (68.3$^{\rm th}$ percentile) equal exactly to 1 $\ms$ with $N_{\rm epochs,\:s2}$ data, independently from the value of $\tau_{\rm AR}$; and \textit{ii}) active stars, $P_{\rm orb}\neq P_{\rm rot}$, long $\tau_{\rm AR}$, and $N_{\rm epochs,\:s2}$ data.  

\subsubsection{Dependence of the results on the choice of the priors}
\label{subsec:alternprior}
\begin{figure}
   \centering
   \includegraphics[width=9cm]{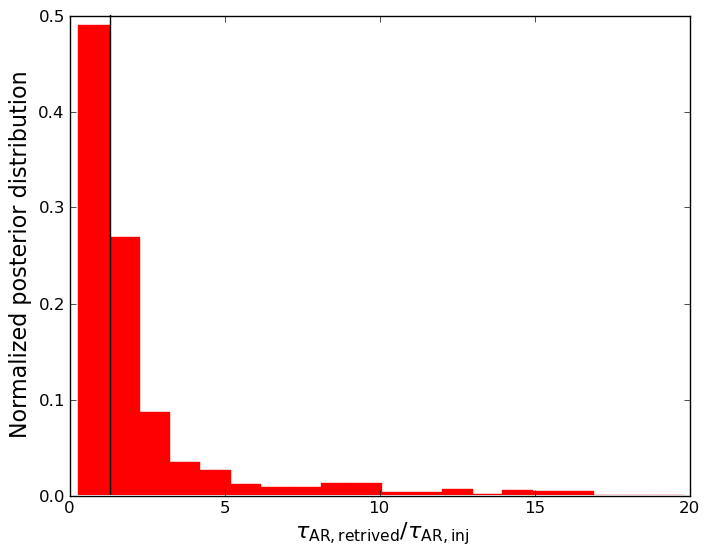}
      \caption{Posterior distribution of the $\tau_{\rm AR,ratio}$=$\frac{\tau_{\rm AR, retrieved}}{\tau_{\rm AR, injected}}$ for the scenario represented by stars with low and quickly variable activity (short $\tau_{\rm AR}$), hosting a planet with $P_{\rm orb}\sim P_{\rm rot}$, and with $N_{\rm epochs,s1}$ data. The vertical lines indicate the median of the distribution. Here we show the result for the alternative priors on $\tau_{\rm AR}$ and $P_{\rm rot}$ (Sect. \ref{subsec:alternprior}), which has to be compared with the upper plot in Fig. \ref{fig:taucomp1}.}
         \label{fig:alternprior1}
   \end{figure}

In our simulations the only parameter we explored over a significantly large scale is $\tau_{\rm AR}$, which we have assumed to be unconstrained from possible ancillary data. As previously discussed (Sect. \ref{sect:results}, \textit{case I}), the hyper-parameter $\tau_{\rm AR}$ is poorly fitted for the case of low-activity stars and $\tau_{\rm AR}\sim P_{\rm orb}$, especially with $N_{\rm epochs, s1}$ data, since the distribution of $\tau_{\rm AR,ratio}$ is bi-modal, as shown in Fig. \ref{fig:taucomp1} (the bi-modality being much less pronounced with $N_{\rm epochs, s2}$ data). In principle this fact may impact the accuracy and precision of the retrieved planetary signal.
Since by adopting that prior we are assuming to be ignorant about the order of magnitude (or scale) of $\tau_{\rm AR}$, a potentially better choice for the prior is adopting an uninformative distribution on $\ln (\tau_{\rm AR})$ within the range [0,6.9] corresponding to the interval [1,1000] days of the linear scale. To investigate the effects of changing the prior, we first analysed with the logarithmic prior 1\,000 mock datasets with $N_{\rm epochs,s1}$ RV measurements concerning the scenario of low-activity stars, short $\tau_{\rm AR}$, and $P_{\rm orb}\neq P_{\rm rot}$. Moreover, we restricted the range of the prior on $P_{\rm orb}$ to $\pm$0.05 days to simulate the case of better known transit ephemeris. From the new analysis we find that $\tau_{\rm AR, ratio}$=1.3$^{\rm +1.7}_{\rm -0.6}$, which represents a significant improvement of the MC fitting analysis for the stellar component (Fig. \ref{fig:alternprior1}). However, for $K_{\rm b, ratio\: 50\%}$ and $K_{\rm b,\:50\%}$/$\sigma^{\rm -}_{\rm K_{\rm b}}$ we do not find any difference with respect to the original set of priors. This shows that in our case a more correct fitting of the stellar signal does not improve the determination of the planetary parameters. 
We get similar results when analysing datasets with $N_{\rm epochs,s2}$ RVs, finding $\tau_{\rm AR, ratio}$=1.1$^{\rm +0.7}_{\rm -0.4}$, i.e. the stellar activity signal is better represented, but this does not come with a better statistical result for the planetary component. 

Using the same set of new priors, we also considered the case with long $\tau_{\rm AR}$. We get $\tau_{\rm AR,ratio}=1.3^{\rm +0.7}_{\rm -0.8}$, which is an improvement in the ability of retrieving the injected $\tau_{\rm AR}$ since the median of $\tau_{\rm AR,ratio}$ is half that found using the original prior. However, even in this case there is no improvement for the parameters $K_{\rm b, ratio\: 50\%}$ and $K_{\rm b,\:50\%}$/$\sigma^{\rm -}_{\rm K_{\rm b}}$.

Finally, as an additional check we considered the case of low-activity stars, short $\tau_{\rm AR}$, and $P_{\rm orb}\sim P_{\rm rot}$, in particular to test the effects of the tighter prior on $P_{\rm orb}$. As expected, we found a significant improvement in fitting the hyper-parameter $\tau_{\rm AR}$, getting $\tau_{\rm AR,ratio}=1.3^{\rm +2.4}_{\rm -0.6}$. For $K_{\rm b, ratio\: 50\%}$ and $K_{\rm b,\:50\%}$/$\sigma^{\rm -}_{\rm K_{\rm b}}$ we do not find any improvement with respect to the values tabulated in Table \ref{Table:mcmcanalysis}.

In conclusion, we find that using a uniform logarithmic prior on $\tau_{\rm AR}$ provides more realistic representations for the quasi-periodic stellar activity signal than adopting a uniform linear prior, but more accurate values for $\tau_{\rm AR}$ does not correspond to more accurate retrieved planetary signals, even restricting the prior on $P_{\rm orb}$. 

\subsubsection{Exploring the effects of orbital periods very close to stellar rotation periods}
Here we focus on the scenarios for which $P_{\rm orb}\sim P_{\rm rot}$ and consider only the datasets for which the injected values satisfy the condition $|P_{\rm orb}-P_{\rm rot}|\le$0.5 days. We choose the range of $\pm$0.5 days because this is the interval we have adopted to define the prior for $P_{\rm orb}$ (Table \ref{Table:simpriors}). In this case, values for $P_{\rm orb}$ are sampled within a parameter space which includes $P_{\rm rot}$. We are interested in the analysis of those subsets to assess possible effects on the retrieved planetary signals due to the very close proximity between the orbital and stellar rotations periods. We note that the range of 0.5 days is 3-4 times smaller than 1$\sigma$ associated to the $P_{\rm orb}-P_{\rm rot}$ input distributions for these scenarios (Table \ref{Table:simuldistr}).
Results are shown in Table \ref{Table:mcmcanalysis} between square brackets. Typically, they are derived from $\sim$20$\%$ of the total samples, a number which allows a good statistics. The only difference we note with respect to the results for the total sample is that K$_{\rm b,ratio\:50\%}$ shifts to higher values by an amount of $\sim50\%$ for the case $\tau_{\rm AR}\gg$P$_{\rm rot}$, both for low- and high-activity stars and independently from the number of observations. This indicates that the fitting procedure tends to absorb part of the stellar signal in the planetary component of the model and, not surprisingly, this happens for stars with the longer evolutionary time scale $\tau_{\rm AR}$, in that the stellar term has properties closer to those of a circular Keplerian. The result suggests that alternative recipes to model the stellar activity term should be considered in addition to GP regression when $P_{\rm orb}$ and $P_{\rm rot}$ are very close.

\subsubsection{Effects of neglecting to model the stellar activity with a GP regression}
Here we consider the case where the stellar activity is fitted by including only the uncorrelated jitter term in the global model, to explore the effects of using a simpler model than the GP regression, as is often done when dealing with real systems. We focus the attention to the scenario corresponding to a quiet star with $P_{\rm rot}\neq P_{\rm orb}$, since fitting the activity with a correlated-noise model (like the GP regression) is expected to be more convenient and effective a-priori for active stars. We analysed a subset of $N$=500 RV datasets for each scenario with $K_{\rm b,\: inj}$=1 \ms (short and long $\tau_{\rm AR}$; $N_{\rm epochs,s1}$ and $N_{\rm epochs,s2}$ RV measurements), and compared the Bayesian evidences of the models with and without the GP term. Table \ref{Table:gpnogpevidence} shows the results for the difference of the logarithm of the Bayesian evidences for the two models, $\Delta$ln$\mathcal{Z}$= ln$\mathcal{Z_{\rm GP+1\:planet}}$-ln$\mathcal{Z_{\rm 1\:planet}}$, and lists the values of $K_{\rm b,\:ratio\:50\%}$ and $K_{\rm b,\: 50\%}/\sigma^{-}_{\rm K_{\rm b}}$ derived from using the model without GP and uncorrelated jitter only. We assume that $\Delta$ln$\mathcal{Z}\geq$5 denotes strong evidence in favour of the GP+1 planet model (e.g. see Table 1 in \citealt{feroz11}). With N=40 measurements the use of the GP regression can be considered generally unnecessary for short values of $\tau_{\rm AR}$, since only $\sim$30$\%$ of the samples are better fitted by a GP+1 planet model. With N=80 measurements the GP+1 planet model provides a better fit in $\sim$60$\%$ of the cases. However, comparing with Table \ref{Table:mcmcanalysis}, we see that $K_{\rm b,\:ratio\:50\%}$ and $K_{\rm b,\: 50\%}/\sigma^{-}_{\rm K_{\rm b}}$ are in agreement with those obtained fitting a GP+1 planet model. For the case of long $\tau_{\rm AR}$, the evidence is by far in favour of the GP+1 planet model, in particular with N=80 measurements, as also confirmed by the fact that a $\sigma_{\rm jit}$ five times larger is required if a GP term is not included in the model. This is expected, since the activity term is more stable over time and easier to characterize. Nonetheless, within the framework of our simulations, even in this case the use of the GP regression does not bring better results than a simpler model in terms of planetary semi-amplitude.

%\item \textit{Notes on the symbols used for the distributions summarized in this Table.} , ,  denote the distributions of the ratio between each retrieved parameter and the corresponding injected value in the mock datasets. 
%    $\sigma_{\rm ratio}$=<$\frac{\sqrt[]{\sigma_{\rm RV}^{\rm 2}(t)+\sigma_{\rm jit}^{2}} }{\sigma_{\rm RV}(t)}$>, for each mock dataset. 
%    
%     K$_{\rm b,ratio\: 68.3\%}$: distribution of ratio between the 68.3$^{\rm th}$ percentile of the planetary semi-amplitude K$_{\rm b}$ retrieved from each mock dataset and the injected $K_{\rm b, injected}$=1 \ms.
%     K$_{\rm b}$/$\sigma^{\rm -}_{\rm K_{\rm b}}$ represents the distribution of the ratio between the best-fit values of K$_{\rm b}$ and the corresponding lower uncertainty, and it quantifies the significance of the retrieved planetary Doppler semi-amplitude with respect to a null value. 

%\begin{landscape}
\begin{table*}
\caption[]{Summary of the main results of the analysis performed with MultiNest (Sect. \ref{sect:results}). The estimates are given as $16^{\rm th}$, $50^{\rm th}$ and $84^{\rm th}$ percentiles of the various distributions we have examined.} 
\begin{threeparttable}
         \label{Table:mcmcanalysis}
   \normalsize
%   \makebox[\textwidth][c]{
   \begin{tabular}{lccccc}
            \hline
            \hline
            \noalign{\smallskip}
             & & \multicolumn{2}{c}{\textbf{P$_{\rm rot}\sim$P$_{\rm orb}$}} & \multicolumn{2}{c}{\textbf{P$_{\rm rot}$ $\neq$ P$_{\rm orb}$} }\\
            \cmidrule(lr){3-4}\cmidrule(lr){5-6}
            \noalign{\smallskip}
           % \hline
            & Parameter & short $\tau_{\rm AR}$ & long $\tau_{\rm AR}$ & short $\tau_{\rm AR}$ & long $\tau_{\rm AR}$ \\
         %   \hline
        %  \cmidrule(lr){3-4}\cmidrule(lr){5-6}
          \noalign{\smallskip}
            \noalign{\smallskip}
            \hline \hline
            \noalign{\smallskip}
            \multicolumn{6}{l}{\textbf{One observing season}} \\ % & & & & & & & & & & & & & \\
            \noalign{\smallskip}
            \hline
            \noalign{\smallskip}
            Low activity star & $h_{\rm ratio}$ &  1.0$\pm$0.4 & 1.1$\pm$0.5 & 1.01$\pm$0.4 & 1.1$^{\rm +0.5}_{\rm -0.4}$ \\
            \noalign{\smallskip}
            & $P_{\rm rot,\: ratio}$ & 1.01$\pm$0.09 & 1.00$\pm$0.02 & 1.0$\pm$0.09 & 1.00$\pm$0.02 \\
            \noalign{\smallskip}
            & $\tau_{\rm AR,\: ratio}$ & 10.1$^{\rm +15.5}_{\rm -8.7}$ & 2.7$^{\rm +0.4}_{\rm -0.7}$ & 9.5$^{\rm +16.3}_{\rm -8.1}$ & 2.7$^{\rm +0.4}_{\rm -0.7}$ \\
            \noalign{\smallskip}
   %         & $\sigma_{\rm jit}$ [\ms] & 1.1$^{\rm +0.9}_{\rm -0.5}$ & 0.6$^{\rm +0.5}_{\rm -0.2}$ & 1.2$^{\rm +0.9}_{\rm -0.6}$ &  0.7$^{\rm +0.5}_{\rm -0.2}$ \\
   %         \noalign{\smallskip}
             & $\sigma_{\rm ratio}$ & 1.2$^{\rm +0.3}_{\rm -0.1}$ & 1.05$^{\rm +0.09}_{\rm -0.02}$ & 1.2$^{\rm +0.4}_{\rm -0.1}$ & 1.06$^{\rm +0.1}_{\rm -0.04}$ \\
            \noalign{\smallskip}
             & K$_{\rm b,\:ratio\: 50\%}$ & 1.2$^{\rm +0.9}_{\rm -0.5}$ & 1.08$^{\rm +0.7}_{\rm -0.5}$ & 0.9$^{\rm +0.7}_{\rm -0.3}$ & 0.8$^{\rm +0.5}_{\rm -0.4}$ \\
             & & [1.2$^{\rm +0.9}_{\rm -0.5}$, 886 samples]\tnote{a} & [1.6$\pm$0.7, 910 samples]\tnote{a} \\
            \noalign{\smallskip}
            & K$_{\rm b,\:ratio\: 68.3\%}$ & 1.6$^{\rm +0.9}_{\rm -0.6}$ & 1.4$^{\rm +0.9}_{\rm -0.6}$ & 1.2$^{\rm +0.7}_{\rm -0.4}$ & 1.1$^{\rm +0.5}_{\rm -0.4}$ \\
            \noalign{\smallskip}
             & K$_{\rm b,\:50\%}$/$\sigma^{\rm -}_{\rm K_{\rm b}}$ & 1.6$^{\rm +0.7}_{\rm -0.1}$ & 1.6$^{\rm +0.6}_{\rm -0.1}$ & 1.60$^{\rm +0.71}_{\rm -0.09}$ & 1.7$^{\rm +0.9}_{\rm -0.2}$ \\
             & & [1.6$^{\rm +0.56}_{\rm -0.09}$]\tnote{a} & [1.6$^{\rm +0.5}_{\rm -0.1}$]\tnote{a} \\
            \noalign{\smallskip}
          %   \cmidrule(lr){2-14}
            Active star & $h_{\rm ratio}$ & 1.0$\pm$0.2 & 1.2$\pm$0.3 & 1.0$\pm$0.2 & 1.3$\pm$0.3 \\
            \noalign{\smallskip}
            & $P_{\rm rot,\: ratio}$ & 1.02$^{\rm +0.12}_{\rm -0.08}$ & 1.000$\pm$0.004 &  1.02$^{\rm +0.12}_{\rm -0.08}$ & 1.00$^{\rm +0.07}_{\rm -0.06}$ \\
            \noalign{\smallskip}
            & $\tau_{\rm AR,\: ratio}$ & 1.3$^{\rm +4.7}_{\rm -0.4}$ & 1.3$^{\rm +0.8}_{\rm -0.3}$ &  1.3$^{\rm +7.3}_{\rm -0.4}$ & 1.3$^{\rm +0.8}_{\rm -0.3}$ \\
            \noalign{\smallskip}
     %       & $\sigma_{\rm jit}$ [\ms] & 3.3$^{\rm +4.9}_{\rm -1.8}$ & 0.8$^{\rm +0.7}_{\rm -0.3}$ & 3.4$^{\rm +4.9}_{\rm -2.0}$ & 0.9$^{\rm +0.7}_{\rm -0.3}$ \\
     %       \noalign{\smallskip}
            & $\sigma_{\rm ratio}$  & 2.2$^{\rm +2.0}_{\rm -1.0}$ & 1.1$^{\rm +0.18}_{\rm -0.05}$ & 2.0$^{\rm +2.1}_{\rm -0.8}$ & 1.11$^{\rm +0.18}_{\rm -0.07}$ \\
            \noalign{\smallskip}
             & K$_{\rm b,\:ratio\: 50\%}$ & 2.14$^{\rm +0.5}_{\rm -0.4}$ & 1.4$^{\rm +0.8}_{\rm -0.7}$ & 1.9$^{\rm +0.6}_{\rm -0.3}$ & 0.9$^{\rm +0.6}_{\rm -0.3}$  \\
             & & [2.2$\pm$0.4, 1395 samples]\tnote{a} & [2.1$^{\rm +0.4}_{\rm -0.3}$, 1360 samples]\tnote{a} \\
             \noalign{\smallskip}
             & K$_{\rm b,\:ratio\: 68.3\%}$ & 3.0$\pm$0.5 & 1.8$^{\rm +1.2}_{\rm -0.8}$ & 2.6$^{\rm +0.6}_{\rm -0.5}$ & 1.2$^{\rm +0.6}_{\rm -0.4}$ \\
            \noalign{\smallskip}
             & K$_{\rm b,\:50\%}$/$\sigma^{\rm -}_{\rm K_{\rm b}}$ & 1.58$^{\rm +0.07}_{\rm -0.05}$ & 1.64$^{\rm +0.3}_{\rm -0.07}$ &  1.6$^{\rm +0.08}_{\rm -0.06}$ & 1.6$^{\rm +0.6}_{\rm -0.1}$ \\
             & & [1.58$^{\rm +0.07}_{\rm -0.05}$]\tnote{a} & [1.54$^{\rm +0.11}_{\rm -0.06}$]\tnote{a} \\
            \noalign{\smallskip}
            \hline
            \noalign{\smallskip}
            \multicolumn{6}{l}{\textbf{Two observing seasons}} \\
            \noalign{\smallskip}
            \hline
            \noalign{\smallskip}
            Low activity star & $h_{\rm ratio}$ & 1.0$_{\rm -0.3}^{\rm +0.2}$ & 1.1$^{\rm +0.4}_{\rm -0.3}$ & 1.0$_{\rm -0.3}^{\rm +0.2}$ & 1.1$\pm$0.3 \\
            \noalign{\smallskip}
            & $P_{\rm rot,\: ratio}$ & 1.00$\pm$0.07 & 1.00$\pm$0.01 & 1.00$_{\rm -0.08}^{\rm +0.09}$ & 1.00$\pm$0.01 \\
            \noalign{\smallskip}
            & $\tau_{\rm AR,\: ratio}$ & 1.5$^{\rm +10.2}_{\rm -0.5}$ & 1.8$^{\rm +1.2}_{\rm -0.2}$ & 1.5$_{\rm -0.5}^{\rm +8.1}$ & 1.7$^{\rm +1.2}_{\rm -0.7}$ \\
            \noalign{\smallskip}
     %       & $\sigma_{\rm jit}$ [\ms] & 0.8$^{\rm +0.9}_{\rm -0.3}$ & 0.5$^{\rm +0.4}_{\rm -0.2}$ & 0.8$^{\rm +0.9}_{\rm -0.4}$  &  0.5$^{\rm +0.4}_{\rm -0.2}$ \\
      %      \noalign{\smallskip}
             & $\sigma_{\rm ratio}$ & 1.07$^{\rm +0.24}_{\rm -0.04}$ & 1.03$^{\rm +0.05}_{\rm -0.02}$ & 1.05$^{\rm +0.12}_{\rm -0.04}$ & 1.04$^{\rm +0.08}_{\rm -0.02}$ \\
            \noalign{\smallskip}
             & K$_{\rm b,\:ratio\: 50\%}$ & 1.0$^{\rm +0.7}_{\rm -0.4}$ & 0.9$^{\rm +0.6}_{\rm -0.4}$ & 0.8$_{\rm -0.3}^{\rm +0.6}$ & 0.8$^{\rm +0.5}_{\rm -0.4}$ \\
             & & [1.1$^{\rm +0.7}_{\rm -0.4}$, 955 samples]\tnote{a} & [1.4$^{\rm +0.8}_{\rm -0.5}$, 917 samples]\tnote{a} \\
             \noalign{\smallskip}
             & K$_{\rm b,\:ratio\: 68.3\%}$ & 1.3$^{\rm +0.7}_{\rm -0.5}$ & 1.2$^{\rm +0.7}_{\rm -0.5}$ & 1.0$^{\rm +0.6}_{\rm -0.3}$ & 1.0$\pm$0.4 \\
            \noalign{\smallskip}
             & K$_{\rm b,\:50\%}$/$\sigma^{\rm -}_{\rm K_{\rm b}}$ & 1.6$^{\rm +0.8}_{\rm -0.1}$ & 1.7$^{\rm +1.2}_{\rm -0.2}$ & 1.6$_{\rm -0.1}^{\rm +0.9}$ & 1.9$^{\rm +1.5}_{\rm -0.4}$ \\
             & & [1.6$^{\rm +0.6}_{\rm -0.1}$]\tnote{a} & [1.6$^{\rm +0.7}_{\rm -0.1}$]\tnote{a} \\
            \noalign{\smallskip}
          %   \cmidrule(lr){2-14}
            Active star & $h_{\rm ratio}$ & 0.9$\pm$0.1 & 1.1$\pm$0.2 & 1.0$\pm$0.1 &  1.1$\pm$0.2 \\
            \noalign{\smallskip}
            & $P_{\rm rot,\: ratio}$ & 1.00$^{\rm +0.05}_{\rm -0.04}$ & 1.000$\pm$0.002 & 1.00$^{\rm +0.05}_{\rm -0.04}$ & 0.999$\pm$0.002 \\
            \noalign{\smallskip}
            & $\tau_{\rm AR,\: ratio}$ & 1.1$^{\rm +0.3}_{\rm -0.2}$ & 1.1$\pm$0.2 & 1.1$^{\rm +0.4}_{\rm -0.2}$ & 1.1$\pm$0.2 \\
            \noalign{\smallskip}
  %          & $\sigma_{\rm jit}$ [\ms] & 2.4$^{\rm +4.1}_{\rm -1.3}$ & 0.6$^{\rm +0.4}_{\rm -0.2}$ & 2.4$^{\rm +4.3}_{\rm -1.2}$ & 0.6$^{\rm +0.5}_{\rm -0.2}$ \\
   %         \noalign{\smallskip}
             & $\sigma_{\rm ratio}$ & 1.7$^{\rm +1.7}_{\rm -0.5}$ & 1.03$^{\rm +0.04}_{\rm -0.02}$ & 1.4$^{\rm +1.6}_{\rm -0.2}$ & 1.07$^{\rm +0.12}_{\rm -0.04}$ \\
            \noalign{\smallskip}
             & K$_{\rm b,\:ratio\: 50\%}$ & 1.9$^{\rm +0.5}_{\rm -0.4}$ & 1.2$^{\rm +0.8}_{\rm -0.7}$ & 1.5$^{\rm +0.7}_{\rm -0.4}$ & 0.8$^{\rm +0.5}_{\rm -0.3}$ \\
             & & [2.0$^{\rm +0.4}_{\rm -0.3}$, 1374 samples]\tnote{a} & [1.9$\pm$0.3, 1368 samples]\tnote{a} \\
             \noalign{\smallskip}
             & K$_{\rm b,\:ratio\: 68.3\%}$ & 2.7$\pm$0.5 & 1.6$^{\rm +1.2}_{\rm -0.8}$ & 2.1$^{\rm +0.8}_{\rm -0.5}$ & 1.0$^{\rm +0.5}_{\rm -0.4}$ \\
            \noalign{\smallskip}
             & K$_{\rm b,\:50\%}$/$\sigma^{\rm -}_{\rm K_{\rm b}}$ & 1.5$^{\rm +0.06}_{\rm -0.05}$ & 1.5$^{\rm +0.5}_{\rm -0.08}$ & 1.54$^{\rm +0.10}_{\rm -0.05}$ & 1.7$^{\rm +1.1}_{\rm -0.2}$ \\
             & & [1.54$^{\rm +0.06}_{\rm -0.04}$]\tnote{a} & [1.50$^{\rm +0.08}_{\rm -0.06}$]\tnote{a}\\
            \noalign{\smallskip}
            \hline       
            \hline
     \end{tabular}  
     \begin{tablenotes}
     \item[a] Derived from posteriors related to simulated datasets that satisfy the condition $|P_{\rm rot}-P_{\rm orb}|\le$0.5 days
   \end{tablenotes}
 %    }
  \end{threeparttable}
\end{table*}
% \end{landscape}

\begin{table}
  \caption[]{Model comparison analysis, with or without a GP quasi-periodic kernel included in the global model. Here we consider the case of quiet stars with $P_{\rm rot}\neq P_{\rm orb}$. The values of $K_{\rm b,\:ratio\:50\%}$ and $K_{\rm b,\: 50\%}/\sigma^{-}_{\rm K_{\rm b}}$ are derived from fitting the model without using GP. $\sigma_{\rm jit,\:no\:GP}$/$\sigma_{\rm jit,\:with\:GP}$ is the ratio between the values of the fitted uncorrelated jitter without and with the GP term included in the model. $\Delta$ln$\mathcal{Z}$= ln$\mathcal{Z_{\rm GP+1\:planet}}$-ln$\mathcal{Z_{\rm 1\:planet}}$}
         \label{Table:gpnogpevidence}
         \centering
         %\small
   \begin{tabular}{lcc}
            \hline
            \hline
            \noalign{\smallskip}
            \textbf{One season of observations} & &\\
            \noalign{\smallskip}
            \hline
            & \textbf{short $\tau_{\rm AR}$} & \textbf{long $\tau_{\rm AR}$}\\
            \noalign{\smallskip}
            $K_{\rm b,\:ratio\:50\%}$ [\ms] & 0.95$^{+0.87}_{-0.35}$ & 0.98$^{+0.68}_{-0.41}$ \\ 
            \noalign{\smallskip}
            $K_{\rm b,\: 50\%}/\sigma^{-}_{\rm K_{\rm b}}$ & 1.6$^{+0.6}_{-0.1}$ & 1.6$^{+0.5}_{-0.1}$ \\
            \noalign{\smallskip}
            median $\sigma_{\rm jit,\:no\:GP}$/$\sigma_{\rm jit,\:with\:GP}$ & 2.3 & 3.6 \\
            \noalign{\smallskip}
            median $\Delta$ln$\mathcal{Z}$ & 2.2 & 5.2 \\
            \noalign{\smallskip}
            number of datasets for which $\Delta$ln$\mathcal{Z}>$5 & 28$\%$ & 51$\%$ \\
            \noalign{\smallskip}
            median $\sigma_{\rm jit,\:no\:GP}$/$\sigma_{\rm jit,\:with\:GP}$ & 2.3 & 3.6 \\
            \noalign{\smallskip}
            \hline
            \noalign{\smallskip}
            \textbf{Two seasons of observations} & &\\
            \noalign{\smallskip}
            \hline
            & \textbf{short $\tau_{\rm AR}$} & \textbf{long $\tau_{\rm AR}$}\\
            \noalign{\smallskip}
            $K_{\rm b,\:ratio\:50\%}$ [\ms] & 0.86$^{+0.58}_{-0.35}$ & 0.74$^{+0.53}_{-0.26}$ \\ 
            \noalign{\smallskip}
            $K_{\rm b,\: 50\%}/\sigma^{-}_{\rm K_{\rm b}}$ & 1.6$^{+0.7}_{-0.1}$ & 1.6$^{+0.5}_{-0.08}$ \\
            \noalign{\smallskip}
            median $\sigma_{\rm jit,\:no\:GP}$/$\sigma_{\rm jit,\:with\:GP}$ & 3.6 & 4.9\\
            \noalign{\smallskip}
            median $\Delta$ln$\mathcal{Z}$ & 6.7 & 20.9 \\
            \noalign{\smallskip}
            number of datasets for which $\Delta$ln$\mathcal{Z}>$5 & 57$\%$ & 85$\%$ \\
            \noalign{\smallskip}
            median $\sigma_{\rm jit,\:no\:GP}$/$\sigma_{\rm jit,\:with\:GP}$ & 3.6 & 4.9 \\
            \noalign{\smallskip}
            \hline
            \hline
     \end{tabular}    
\end{table}
%------------------------------------------------------------------------------------------

\section{Periodogram analysis}
\label{sect:periodanalysis}

Here we present the results of a blind statistical analysis performed on a large sub-sample of the simulated datasets, using different periodogram algorithms commonly used to investigate the frequency content of RV time series. The goal of this Section is to examine and characterize some statistical properties, described in Table \ref{Table:analysedparamperiodogram}, of periodograms for RV datasets containing quasi-periodic stellar activity signals.

\begin{table*}
  \caption[]{Description of quantities analysed in this work to characterize the performances of the GLS, BGLS and FREDEC algorithms to search for periodic signals in the mock RV datasets (Sect. \ref{sect:periodanalysis}, Tables \ref{Table:periodanalysis1} and \ref{Table:periodanalysis2}).}
         \label{Table:analysedparamperiodogram}
         \centering
         \scriptsize
   \begin{tabular}{lp{0.7\textwidth}}
            \hline
            \hline
            \noalign{\smallskip}
            \textbf{Quantity}  & \textbf{Description} \\
            \noalign{\smallskip}
            $N_{\rm FAP<1\%}$, $N_{\rm FAP\: 1-10\%}$ & Number of solutions detected with FAP<1$\%$ and 1$\%$<FAP<10$\%$ respectively, relative to a total of 1\,000 examined periodograms.  \\ 
            \noalign{\smallskip}
        %    $P_{\rm rot,ratio}$ ($f_{\rm rot,ratio}$) & $\frac{P_{\rm out}}{P_{\rm rot}}$ ($\frac{f_{\rm out}}{f_{\rm rot, inj}}$), ratio between the retrieved period (frequency) and the highest peak in the periodogram \\ 
        %    & and the corresponding injected stellar rotation period (frequency). \\
        %    \noalign{\smallskip}
        %    $P_{\rm orb,ratio}$ ($f_{\rm orb,ratio}$) & $\frac{P_{\rm out}}{P_{\rm orb,inj}}$ ($\frac{f_{\rm out}}{f_{\rm orb,inj}}$), ratio between the retrieved period (frequency) and the highest peak in the periodogram \\
        %     & and the corresponding injected planetary orbital period (frequency).\\
        %    \noalign{\smallskip}
             $C_{P_{\rm rot}}\equiv$ completeness $P_{\rm rot}$ [$\%$] & Ratio between the number of datasets with the highest peak in the periodograms \textit{i)} with FAP<1$\%$, and \textit{ii)} within $\pm$10$\%$ of the injected $P_{\rm rot}$, and the total number of the analysed datasets. The same quantity is calculated for the first harmonic $P_{\rm rot}$/2.\\
            \noalign{\smallskip}
             $C_{P_{\rm orb}}\equiv$ completeness $P_{\rm orb}$ [$\%$] & Same as for $C_{P_{\rm rot}}$, but for the injected orbital period $P_{\rm orb}$.\\
             $R\equiv$ reliability [$\%$] & Ratio between the number of datasets with the highest peak in the periodograms \textit{i)} with FAP<1$\%$, and \textit{ii)} within $\pm$10$\%$ of the injected $P_{\rm rot}$, and the number of analysed datasets with highest peaks in the periodograms having FAP<1$\%$. We calculated the same ratio for the first harmonic of the stellar rotation period $P_{\rm rot/2}$ (indicated between brackets in Table \ref{Table:periodanalysis1} and \ref{Table:periodanalysis2}).\\
            % R$_{P_{\rm orb}}\equiv$reliability $P_{\rm orb}$ [$\%$] & Same as for \textit{reliability} $P_{\rm rot}$, but for the injected orbital period $P_{\rm orb}$ of the planet.\\
            \noalign{\smallskip}
            \hline
            \hline
     \end{tabular}    
\end{table*}

As done by \cite{pinamontietal2017}, the datasets were analysed with softwares that calculate the frequency content in timeseries: the Generalized Lomb-Scargle periodograms (GLS, \citealt{zech09}); the Bayesian Generalised Lomb-Scargle (BGLS, \citealt{mortieretal2015}); and the FREquency DEComposer (FREDEC, \citealt{baluev2013}).
GLS performs a simple sinusoidal fit of the data, weighting them on the measurements errors. BGLS generalizes this approach taking into account Bayesian probability. FREDEC performs a multi-frequency analysis of the data, decomposing a time series into a number of sinusoids. The algorithm first selects candidate frequencies, then computes the statistical significance for each frequency tuple, providing as solution the most significant multi-frequency combination. We refer the reader to the indicated references for more details about each algorithm.

Contrary to the analysis described in Section \ref{sect:results}, here we assume that nothing is known a-priori about the signal content in the RV time series. This part of the work should be considered an extension of the comparative analysis discussed in \cite{pinamontietal2017}, in that here we analysed more thoroughly the performances of the algorithms on datasets containing a realistic description of stellar activity RV signals.
To asses the detectability of the simulated planetary signals, we can adopt the $K/N$ ratio, defined as in \citet{dumusque17}:
\begin{equation}
K/N = {K_{\rm b} \over \text{RV}_\text{RMS}} \times \sqrt[]{N_\text{obs}},
\end{equation}
where $N_\text{obs}=40,80$ are the number of observations for the one and two seasons cases, respectively. We see from Table \ref{Table:simuldistr} that $K/N$ is generally well below the $K/N$=7.5 threshold proposed by \cite{dumusque17}, making planet detection a very challenging task.   

The GLS periodograms are computed using the code implementation in the \texttt{astropy2.0.2} library for \texttt{python3.5}. The False Alarm Probabilities (FAP), or \textit{p}-values, are computed with a bootstrap with replacement analysis using 1\,0000 RV mock datasets.
The BGLS periodograms have been calculated with an \texttt{IDL} version of the publicly available \texttt{python} code\footnote{https://www.astro.up.pt/exoearths/tools.html}, developed for the analysis discussed in \citet{pinamontietal2017}. The FAPs for the Bayesian periodograms are computed as described in \citet{pinamontietal2017}. The analysis with FREDEC has been carried out with the publicly available \texttt{C++} version of the code\footnote{https://sourceforge.net/projects/fredec/}. 
The FAPs associated to the output frequency tuples are computed with the built-in function of the code.

We computed periodograms for subsets composed of 1\,000 mock RV datasets, one for each of the 16 different simulated scenarios described in Sect. \ref{sect:simsetup}.
To ease the automatic identification of the main periodicities in the time series, we calculated the periodograms exploring periods longer than 5 days. This cut helps avoiding the daily aliases in the spectral window, which would be discarded in a real single-target analysis. Moreover, this threshold should not affect the retrieval of the injected signals, since no values of $P_{\rm rot}\leq$5 d or $P_{\rm orb}\leq$5 d have been simulated. 
For the sake of robustness of period detection, we set the upper limit of the period search interval to half the timespan of the data, in order to ensure at least two full cycles for every test frequency.% Choosing this range also avoids the long periods artificially returned by the periodogram analyses when a long term trend (either stochastic or astrophysical) is present in the data. 

%\textcolor{red}{Idee maturate dalla discussione con Matteo e Gaetano:
% Investigare i risultati in termini di FAP calcolata in automatico dai codici: scartare i risultati con FAP > 10$\%$ (ma quantificare quanti sono); suddividere i risultati per bin di FAP: <1$\%$ e tra 1-10$\%$ (e includere il caso 10-50$\%$ solo per GLS/BGLS). Questa analisi e utile per capire che peso dare ai valori di FAP forniti in automatico: nel nostro caso sappiamo cosa abbiamo simulato e quindi possiamo "controllare" la situazione
% Con "risultati" si intendono istogrammi con distribuzione dei rapporti P$_{\rm out}$/P$_{\rm rot,inj}$ e P$_{\rm out}$/P$_{\rm orb,inj}$
%- Investigare piu in dettaglio il caso di stella quieta e con dati per due stagioni. In questo caso il segnale di attività e poco piu elevato del segnale planetario, e si puo valutare che differenze ci sono tra i casi P$_{\rm orb,inj}\sim$P$_{\rm rot,inj}$ e P$_{\rm orb,inj}\neq$P$_{\rm rot,inj}$, per cercare di rispondere in modo quantitativo alle domande "quanto interferisce nel risultato finale il fatto che P$_{\rm orb,inj}$ e P$_{\rm rot,inj}$ non siano troppo diversi?" e "che ruolo ha un differente tempo scala $\lambda$?"}

We defined the parameters described in Table \ref{Table:analysedparamperiodogram} as figure of merit for assessing the performances of the algorithms. In the case of FREDEC, we considered for the purpose of computing the completeness $C$ and reliability $R$ all the significant solution tuples with at least one of the output periods within $10\%$ of the input period, even in the presence of additional output periodicities. In the following, we present some major outcomes of the analysis, selected from the complete list of results shown in Tables \ref{Table:periodanalysis1} and \ref{Table:periodanalysis2}. First, we organize the discussion into subgroups according to the different simulated scenarios (e.g. different stellar activity levels), then focusing on the comparison among the algorithms. This way of presenting should allow for an easier identification, and therefore better exploitation of the results. 
	
We note that for all the cases with $P_{\rm rot}\sim P_{\rm orb}$ we do not report the completeness $C_{P_{\rm orb}}$ for the orbital period. In fact, we do not consider this as an informative quantity for our purposes, since the input periods $P_{\rm rot}$ and $P_{\rm orb}$ used to build the datasets are too close to be disentangled by a periodogram analysis. 

\subsection{Dependence of the algorithm performances from the stellar activity properties}

\subparagraph{Case I. Low-activity stars.} In general, the reliability $R$ decreases (or does not improve) passing from $N_{\rm epochs, s1}$ to $N_{\rm epochs, s2}$ data, independently from the values of $\tau_{\rm AR}$, except for BGLS (cases with long $\tau_{\rm AR}$). Instead the fraction of significant signals $N_{\rm FAP<1\%}$ increases considerably with two semesters of measurements, and this implies that the algorithms find a larger number of false positives when more RVs are available. In general, the completeness for $P_{\rm orb}$ ($C_{P_{\rm orb}}$) and the reliability are higher for the cases with long $\tau_{\rm AR}$, both for $P_{\rm rot}\sim P_{\rm orb}$ and $P_{\rm rot}\neq P_{\rm orb}$. This is not surprising, in that a quasi-periodic activity signal with long $\tau_{\rm AR}$ is expected to approach a periodic signal over the time span considered here, therefore a sinusoid (or a combination of sinusoids) can fit the data more efficiently.

\subparagraph{Case II. Active stars.} As for \textit{case I}, we note that in general the performances, in terms of completeness and reliability, are higher when $\tau_{\rm AR}$ is longer. In absolute terms, this behavior is more clear than for low-activity stars with $N_{\rm epochs, s1}$ data, where $R$ is high and close to 100$\%$ for GLS and FREDEC and >90$\%$ for BGLS, including the term for the first harmomic of $P_{\rm rot}$ the calculation. This is not surprising, since the stellar activity signal is stronger and more influenced by the value of $\tau_{\rm AR}$. Passing from short to long $\tau_{\rm AR}$, a great relative improvement in detecting correct signals is also observed for datasets with $N_{\rm epochs, s2}$ RVs, which is higher than that observed for low-activity stars.

\begin{figure}
\centering
\includegraphics[width=9cm]{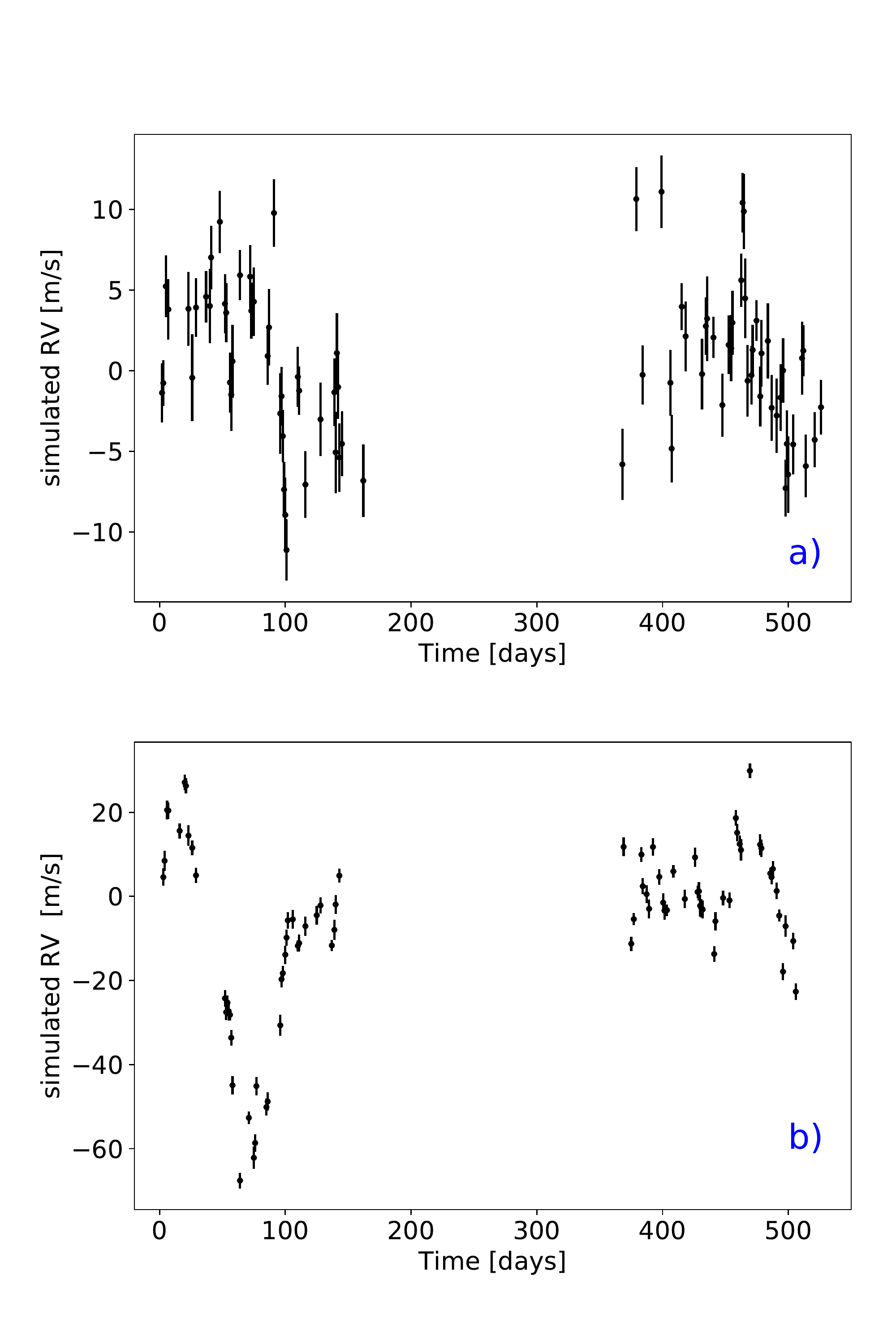}
\caption{\label{fig:diversiquietataushort} Examples of mock RV datasets spanning two seasons for which the injected active regions evolutionary time scale $\tau_{\rm AR}$ is close to the stellar rotation period. (Plot \textit{a}: Low-activity star with $P_{\rm rot}\ne P_{\rm orb}$. Plot \textit{b}: active star with $P_{\rm rot}\sim P_{\rm orb}$). When searched for periodicities using algorithms that fit a sinusoid or a linear combination of sinusoids, the periodograms of such datasets show the main peak ranging between $\sim$10-20 times $P_{\rm rot}$.}
\end{figure}

\subparagraph{Case III. Low-activity vs. active stars for short $\tau_{\rm AR}$ values.} When the stellar activity signal is rapidly evolving, the reliability $R$ is higher for quiet stars, independently from the number of observations. One possible explanation is that the simulated $P_{\rm rot}$ are shorter for active stars (Table \ref{Table:simsetup}) and, the sampling being similar on average for all the cases, it is more difficult modelling strong and rapidly variable signals, taking into account that $\tau_{\rm AR}$ is even shorter for active stars by construction. 
Moreover, we note that for active stars a high percentage of datasets with $N_{\rm epochs, s2}$ RVs, patterns are generated that are mistaken for long-term modulations when fitting the data with single sinusoids, coupled with a larger range of test periods (Fig. \ref{fig:diversiquietataushort}). This is the main reason why the reliability is low. A lower number of spurious long-period signals is found in simulated datsets with $N_{\rm epochs, s1}$ data, but this is also due to the shorter range of test periods explored by the algorithms. Analogously to the active star case, $R$ decreases with increasing number of RVs in low-activity stars, again due to an artificial long-term modulation introduced in the data. Fig. \ref{fig:diversiquietataushort} shows two examples of simulated datasets for stars with low and high activity levels, for which the algorithms return a peak at a period much larger than the injected $P_{\rm rot}$.

\subparagraph{Case IV. Orbital period completeness $C_{P_{\rm orb}}$ for datasets with $P_{\rm rot}\neq P_{\rm orb}$.} The results in Table \ref{Table:periodanalysis2} show that the $C_{P_{\rm orb}}$ is in general less than 3$\%$, and lower than $C_{P_{\rm rot}}$ (much lower for stars with long $\tau_{\rm AR}$). For the case of low-activity stars with long $\tau_{\rm AR}$ there is a slight improvement passing from $N_{\rm epochs, s1}$ to $N_{\rm epochs, s2}$ data, since $P_{\rm orb}$<$P_{\rm rot}$ for construction, and one could expect the detection in the periodogram to be eased. However, the percentages of $C_{P_{\rm orb}}$ stay low, and this denote that a simple frequency analysis of the RVs does not allow to infer the existence of the planetary signal. For low-activity stars, the combination of the sampling and internal precision used in our work can prevent higher values for $C_{P_{\rm orb}}$.

\begin{figure}
\centering
\includegraphics[width=9cm]{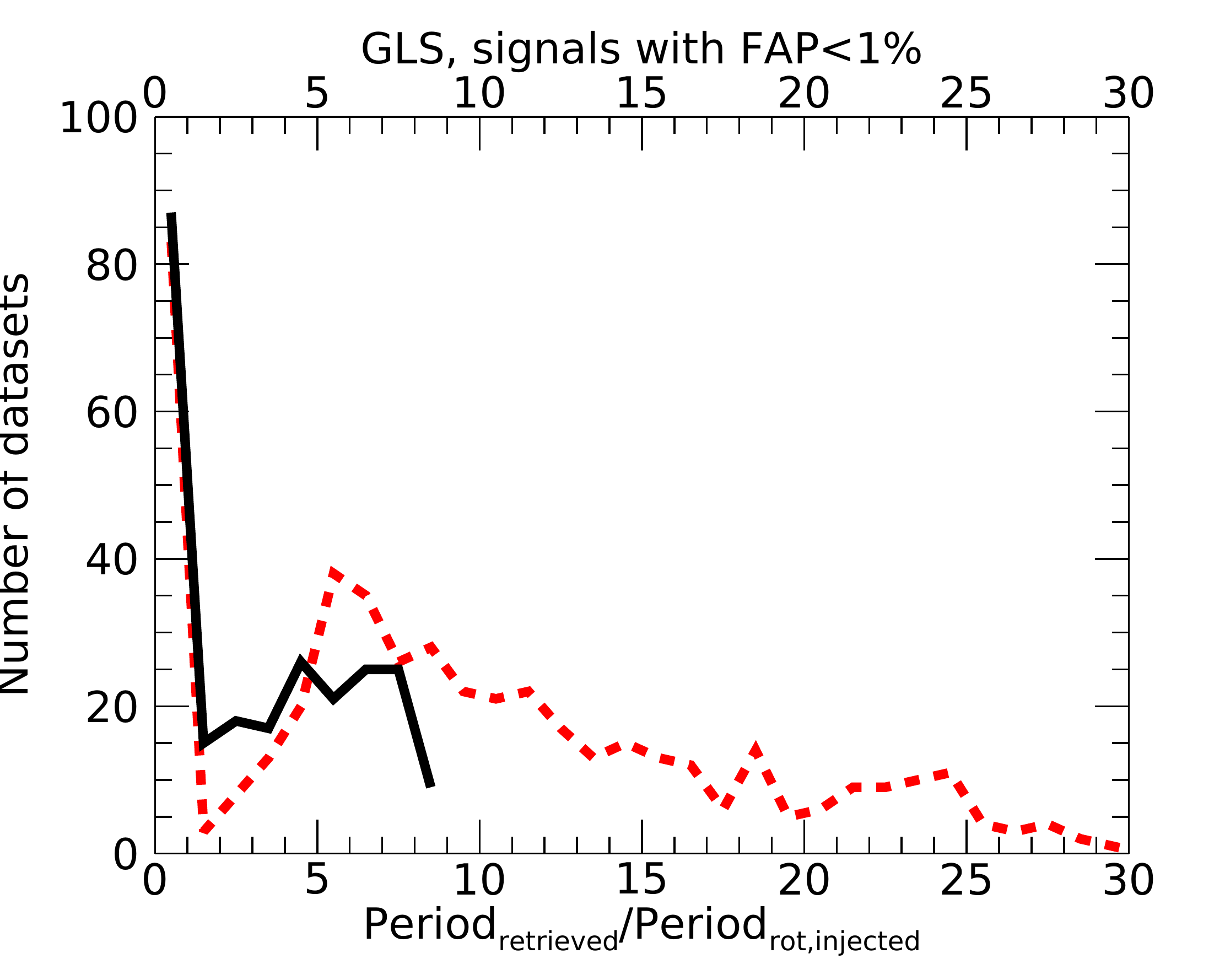}
\includegraphics[width=9cm]{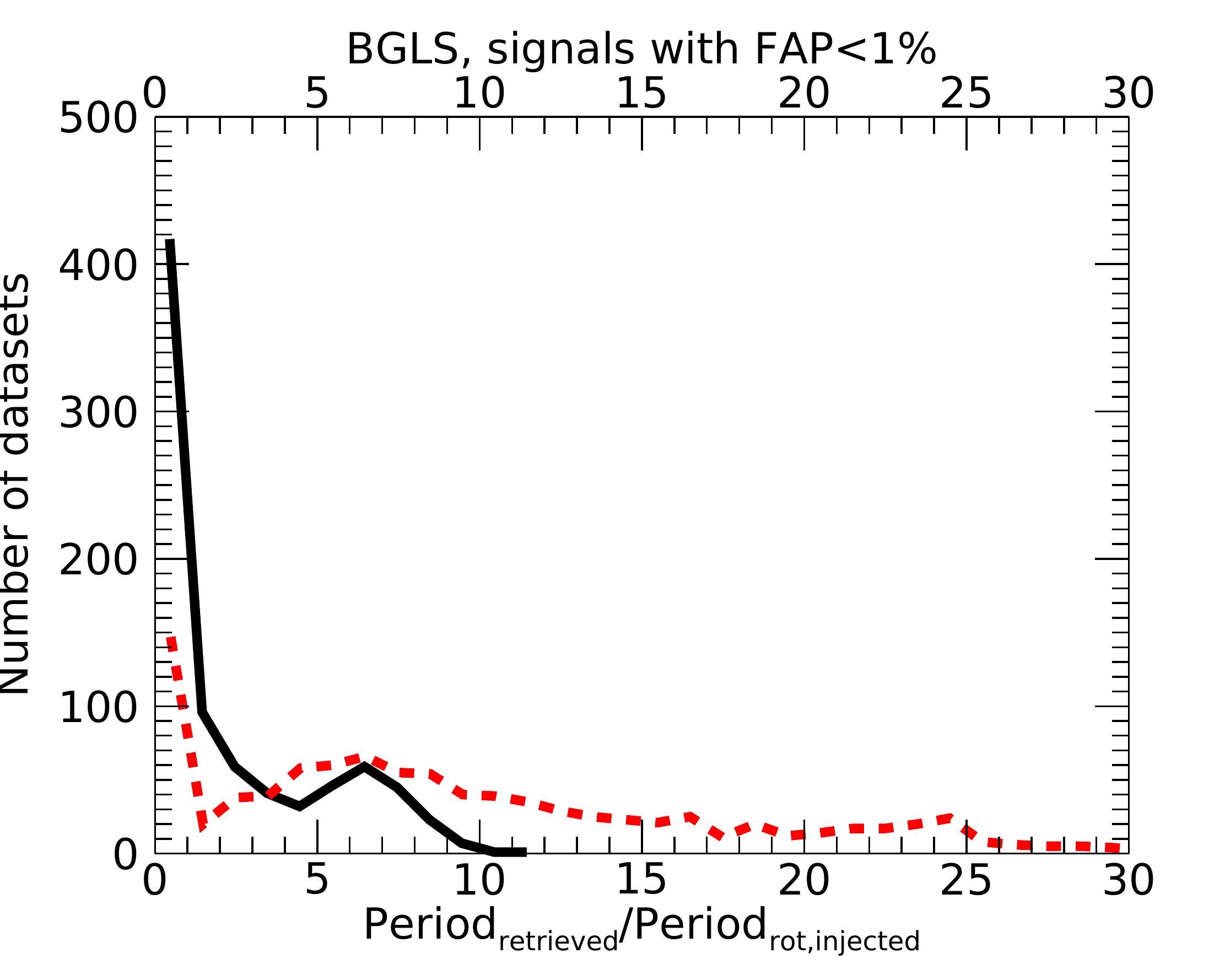}
\caption{\label{fig:glsperdiversiattivataushortsea2} Distributions of the retrieved/simulated period ratios (the injected periods correspond to $P_{\rm rot}$) for the cases of active stars, $P_{\rm rot}\ne$P$_{\rm orb}$, short $\tau_{\rm AR}$, N$_{\rm epochs, s1}$ (solid black line) and $N_{\rm epochs, s2}$ (dashed red line) data. Here we show the results for GLS and BGLS algorithms.}
\end{figure}  

\subsection{Direct comparison of the algorithm performances}
%\subparagraph{Completeness and reliability of the retrieved peaks.} 
In all the cases discussed in this work, we note that the reliability $R$ of GLS decreases passing from $N_{\rm epochs, s1}$ to $N_{\rm epochs, s2}$ data. This is a consequence of the increased number of significant peaks $N_{\rm FAP<1\%}$ with increasing RV measurements. For low-activity stars, the decrease in $R$ is accompanied by an increase in the completeness $C_{\rm P_{\rm rot}}$ with an increasing number of RV data, i.e. GLS retrieves $P_{\rm rot}$ easier despite it results to be less reliable.
In almost all the cases discussed in this work, FREDEC algorithm performs more reliably than GLS and BGLS, as shown by the values of the $R$ parameter, i.e. it is less affected by false positive signals. This is in line with the results of \cite{pinamontietal2017}. Opposite to GLS, $R$ does not change appreciably passing from $N_{\rm epochs, s1}$ to $N_{\rm epochs, s2}$ measurements. For those cases where $N_{\rm FAP<1\%}$ increases (e.g. low-active stars with $P_{\rm rot}\sim$P$_{\rm orb}$), this translates to an increased number of correct detections. In all the cases of active stars with long $\tau_{\rm AR}$, the reliability $R$ is close to 100$\%$, even without including those cases where only the first harmonic $P_{\rm rot}$/2 is recovered. For active stars with short $\tau_{\rm AR}$, $R$ is significantly higher for FREDEC, showing that this algorithm is more effective with quasi-periodic signals variable over short time scales, that are difficult to model with single sinusoids, as done by GLS and BGLS. 

Concerning BGLS, we note that for low-activity stars and $N_{\rm epochs, s1}$ measurements, $C_{\rm P_{\rm rot}}$ and $R$ are much lower than for GLS and FREDEC, indicating lower performance in retrieving the stellar signal, both for short and long timescale $\tau_{\rm AR}$. At the same time, BGLS is particularly sensitive to the first harmonic $P_{\rm rot}$/2, which is detected more often with respect to the number of significant signals $N_{\rm FAP<1\%}$, that is less than the corresponding for GLS and FREDEC. This is due to the definition of the FAP for the BGLS algorithm, which is based on the relative probabilities of the peaks in a periodogram \citep{pinamontietal2017}. When less data are available, several signals are retrieved with similar probabilities, determining high FAP values. This feature is relaxed for $N_{\rm epochs, s2}$ data.

For active stars in general, BGLS finds more significant signals ($N_{\rm FAP<1\%}$ is much higher than for GLS and FREDEC), but the reliability $R$ is comparable to, or even lower than that of the other algorithms. This implies that BGLS finds more false positive signals in these cases.

\begin{table*}
\caption[]{Summary of the main results of the analysis performed on the periodograms calculated with GLS, BGLS and FREDEC (Sect. \ref{sect:periodanalysis}) for the cases where $P_{\rm rot}\sim P_{\rm orb}$. The statistics concerns the peaks detected with a FAP<1$\%$. The results are given as $16^{\rm th}$, $50^{\rm th}$ and $84^{\rm th}$ percentiles of the parameter distributions.} 
\begin{threeparttable}
         \label{Table:periodanalysis1}
   \scriptsize
%   \makebox[\textwidth][c]{
   \begin{tabular}{lccccccc}
            \hline
            \hline
         %   \noalign{\smallskip}
             & & \multicolumn{6}{c}{\textbf{$P_{\rm rot}\sim P_{\rm orb}$}}\\
            \cmidrule(lr){3-8}
          %  \noalign{\smallskip}
           % \hline
            & & \multicolumn{3}{c}{short $\tau_{\rm AR}$} & \multicolumn{3}{c}{long $\tau_{\rm AR}$} \\
         %   \hline
          \cmidrule(lr){3-5}\cmidrule(lr){6-8}
          % \noalign{\smallskip}
            & Parameter & GLS\tnote{1} & BGLS\tnote{1} & FREDEC\tnote{2} & GLS\tnote{a} & BGLS\tnote{a} & FREDEC\tnote{2}  \\
           % \noalign{\smallskip}
            \hline \hline
           % \noalign{\smallskip}
            \multicolumn{8}{l}{\textbf{One observing season}} \\ % & & & & & & & & & & & & & \\
           % \noalign{\smallskip}
            \hline
           % \noalign{\smallskip}
            Low activity star & $N_{\rm FAP<1\%}$ & 24.0$\%$ & 7.1$\%$ & 27.1$\%$ & 50.5$\%$ & 9.0$\%$ & 47.1$\%$  \\
            \noalign{\smallskip}
            & $N_{\rm FAP\: 1-10\%}$ & 33.1$\%$ & 17.7$\%$ & 23.0$\%$ & 21.2$\%$ & 16.6$\%$ & 18.0$\%$ \\
            \noalign{\smallskip}
            & $C_{P_{\rm rot}}$ ($C_{P_{\rm rot}/2}$)\tnote{3}\: [$\%$] & 12.4 (1.0) & 2.0 (1.6) & 12.8 (1.3) & 42.5 (6.1) & 4.1 (3.1) & 40.4 (4.1)\\
            \noalign{\smallskip}
            & $R$ ($+{P_{\rm rot}/2}$)\tnote{3}\: [$\%$] & 51.5 (+4.0) & 28.4 (+22.4) & 54.6 (+4.4) & 84.1 (+12.1) & 45.9 (+35.3) & 88.1 (+8.7)  \\
            \noalign{\smallskip}
             \cmidrule(lr){2-8}
            Active star & $N_{\rm FAP<1\%}$ & 23.0$\%$ & 83.2$\%$ & 27.9$\%$ & 72.8$\%$ & 90.3$\%$ & 66.6$\%$ \\
            \noalign{\smallskip}
            & $N_{\rm FAP\: 1-10\%}$ & 40.0$\%$ & 7.9$\%$ & 30.7$\%$ & 11.5$\%$ & 4.9$\%$ & 15.6$\%$ \\
            \noalign{\smallskip}
            & $C_{P_\text{rot}}$ ($C_{P_{\rm rot}/2}$)\tnote{3}\: [$\%$] & 5.1 (0.4) & 22.2 (3.3) & 8.8 (0.6) & 66.5 (6) & 76.4 (7.2) & 64.8 (1.5) \\
            \noalign{\smallskip}
            & $R$ ($+{P_{\rm rot}/2}$)\tnote{3}\: [$\%$] & 22.3 (+1.7) & 26.7 (+4.0) & 37.6 (+1.8) & 91.3 (+8.3) &  84.6 (+8) & 97.3 (+2.2) \\
            \noalign{\smallskip}
            \hline
           % \noalign{\smallskip}
            \multicolumn{8}{l}{\textbf{Two observing seasons}} \\
           % \noalign{\smallskip}
            \hline
         %   \noalign{\smallskip}
            Low activity star & $N_{\rm FAP<1\%}$ & 47.9$\%$ & 45.7$\%$ & 36.5$\%$ & 74.4$\%$ & 58.2$\%$ & 56.0$\%$ \\
            \noalign{\smallskip}
            & $N_{\rm FAP\: 1-10\%}$ & 31.7$\%$ & 20.2$\%$ & 35.0$\%$ & 16.2$\%$ & 16.9$\%$ & 30.9$\%$ \\
            \noalign{\smallskip}
            & $C_{P_\text{rot}}$ ($C_{P_{\rm rot}/2}$)\tnote{3}\: [$\%$] & 14.6 (1.1) & 10.3 (1.5) & 15.8 (1.5) & 54.7 (5.5) & 37.1 (3.4) & 49.1 (2.5) \\
            \noalign{\smallskip}
            & $R$ ($+{P_{\rm rot}/2}$)\tnote{3}\: [$\%$] & 30.5 (+2.2) & 22.5 (+3.3) & 52.6 (+3.8) & 73.5 (+7.3) & 63.7 (+5.9) & 88.0 (+4.5) \\
            \noalign{\smallskip}
             \cmidrule(lr){2-8}
            Active star & $N_{\rm FAP<1\%}$ & 45.8$\%$ & 95.9$\%$ & 33.5$\%$ & 89.4$\%$ & 97.7$\%$ & 65.8$\%$ \\
            \noalign{\smallskip}
            & $N_{\rm FAP\: 1-10\%}$ & 34.6$\%$ & 1.9$\%$ & 42.7$\%$ & 5.6$\%$ & 1.1$\%$ & 28.9$\%$ \\
            \noalign{\smallskip}
            & $C_{P_\text{rot}}$ ($C_{P_{\rm rot}/2}$)\tnote{3}\: [$\%$] & 5.1 (0.4) & 8.9 (1.3) & 12.9 (0.6) & 60.2 (4.5) & 49.2 (3.9) & 63.6 (1.1) \\
            \noalign{\smallskip}
            & $R$ ($+{P_{\rm rot}/2}$)\tnote{3}\: [$\%$] & 11.2 (+0.9) & 9.3 (+1.4) & 43.9 (+1.5) & 67.4 (+5.1) & 50.4 (+4.0) & 96.7 (+1.7) \\
            \noalign{\smallskip}
            \hline       
            \hline
     \end{tabular}  
     \begin{tablenotes}
     \item[1] The false alarm probability (FAP) has been evaluated through a bootstrap with replacement.
     \item[2] The false alarm probability (FAP) is the one provided by the code.
     \item[3] The percentage in parenthesis is related to the first harmonic of $P_{\rm rot}$. For FREDEC, this percentage corresponds to the number of datasets for which only $P_{\rm rot/2}$ is found as significant peak in the periodogram, while $P_{\rm rot}$ is not.
   \end{tablenotes}
 %    }
  \end{threeparttable}
\end{table*}

\begin{table*}
\caption[]{Summary of the main results of the analysis performed on the periodograms obtained with GLS, BGLS and FREDEC (Sect. \ref{sect:periodanalysis}) for the case where $P_{\rm rot}$ $\neq$ $P_{\rm orb}$. The statistics concern the peaks detected with a FAP<1$\%$. The results are given as $16^{\rm th}$, $50^{\rm th}$ and $84^{\rm th}$ percentiles of the parameter distributions.} 
\begin{threeparttable}
         \label{Table:periodanalysis2}
   \scriptsize
%   \makebox[\textwidth][c]{
   \begin{tabular}{lccccccc}
            \hline
            \hline
         %   \noalign{\smallskip}
             & & \multicolumn{6}{c}{\textbf{$P_{\rm rot}$ $\neq$ $P_{\rm orb}$} }\\
            \cmidrule(lr){3-8}
          %  \noalign{\smallskip}
           % \hline
            &&\multicolumn{3}{c}{short $\tau_{\rm AR}$} & \multicolumn{3}{c}{long $\tau_{\rm AR}$}\\
         %   \hline
          \cmidrule(lr){3-5}\cmidrule(lr){6-8}
          % \noalign{\smallskip}
            & Parameter & GLS\tnote{1} & BGLS\tnote{1} & FREDEC\tnote{2} & GLS\tnote{a} & BGLS\tnote{a} & FREDEC\tnote{2}  \\
           % \noalign{\smallskip}
            \hline \hline
           % \noalign{\smallskip}
            \multicolumn{8}{l}{\textbf{One observing season}} \\ % & & & & & & & & & & & & & \\
           % \noalign{\smallskip}
            \hline
           % \noalign{\smallskip}
            Low activity star & $N_{\rm FAP<1\%}$ & 19.1$\%$ & 8.5$\%$ & 23.0$\%$ & 42.4$\%$ & 10.9$\%$ & 38.5$\%$ \\
            \noalign{\smallskip}
            & $N_{\rm FAP\: 1-10\%}$ & 35.5$\%$ & 18.6$\%$ & 25.6$\%$ & 26.7$\%$ & 17.8$\%$ & 23.2$\%$ \\
            \noalign{\smallskip}
            &  $C_{P_{\rm rot}}$ ($C_{P_{\rm rot}/2}$)\tnote{3}\: [$\%$] & 7.8 (2.7) & 1.2 (2.2) & 8.7 (2.5) & 33.9 (6.3) & 5.2 (3.2) & 32.1 (4.6) \\
            \noalign{\smallskip}
            &  $C_{P_{\rm orb}}$ [$\%$] & 1.6 & 1.4 & 2.3 & 1.6 & 0.9 & 2.3 \\
            \noalign{\smallskip}
            & $R$ ($+{P_{\rm rot}/2}$)\tnote{3}\: [$\%$] & 49.2 (+14.1) & 30.4 (+25.3) & 44.8 (+6.5) & 83.8 (+14.9) & 54.9 (+29.4) & 86.2 (+10.4) \\
            %\noalign{\smallskip}
            %&  $R_{P_{\rm orb}}$ [$\%$] & 8.5 & 16.5 & 7.3 & 3.8 & 7.8 & 5.2 \\
            \noalign{\smallskip}
             \cmidrule(lr){2-8}
            Active star & N$_{\rm FAP<1\%}$ & 24.8$\%$ & 84.3$\%$ & 31.8$\%$ & 69.7$\%$ & 90.3$\%$ & 64.2$\%$ \\
            \noalign{\smallskip}
            & N$_{\rm FAP\: 1-10\%}$ & 36.0$\%$ & 8.6$\%$ & 29.8$\%$ & 15.2$\%$ & 4.9$\%$ & 17.7$\%$ \\
            \noalign{\smallskip}
            & $C_{P_\text{rot}}$ ($C_{P_{\rm rot}/2}$)\tnote{3}\: [$\%$] & 5.6 (0.3) & 18.1 (2.7) & 8.7 (0.5) & 62.3 (6.5) & 74.0 (7.9) & 60.8 (2.6) \\
            \noalign{\smallskip}
            & $C_{P_\text{orb}}$ [$\%$] & 0.8 & 3.0 & 1.3 & 0.5 & 2.7 & 1.4 \\
            \noalign{\smallskip}
            & $R$ ($+{P_{\rm rot}/2}$)\tnote{3}\: [$\%$] & 25.9 (+1.2) & 25.0 (+3.1) & 31.1 (+1.6) & 90.0 (+9.3) & 84.9 (+8.8) & 95.6 (+4.0) \\
            %\noalign{\smallskip}
            %& $R_{P_\text{orb}}$ [$\%$] & 3.3 & 3.5 & 2.6 & 0.7 & 3.0 & 1.7 \\
            \noalign{\smallskip}
            \hline
           % \noalign{\smallskip}
            \multicolumn{8}{l}{\textbf{Two observing seasons}} \\
           % \noalign{\smallskip}
            \hline
         %   \noalign{\smallskip}
            Low activity star & N$_{\rm FAP<1\%}$ & 46.9$\%$ & 45.7$\%$ & 38.1$\%$ & 71.6$\%$ & 57.8$\%$ & 54.7$\%$ \\
            \noalign{\smallskip}
            & N$_{\rm FAP\: 1-10\%}$ & 30.9.4$\%$ & 20.4$\%$ & 33.4$\%$ & 18.2$\%$ & 16.0$\%$ & 30.5$\%$ \\
            \noalign{\smallskip}
            & $C_{P_\text{rot}}$ ($C_{P_{\rm rot}/2}$)\tnote{3}\: [$\%$] & 8.2 (0.8) & 5.7 (1.1) & 12.5 (1.1) & 52 (7.1) & 38.6 (3.7) & 46.2 (3.2) \\
            \noalign{\smallskip}
            & $C_{P_\text{orb}}$\: [$\%$] & 1.3 & 0.9 & 2.9 & 2.2 & 1.2 & 5.2 \\
            \noalign{\smallskip}
            & $R$ ($+{P_{\rm rot}/2}$)\tnote{3}\: [$\%$] & 20.2 (+1.6) & 14.3 (+2.4) & 38.1 (+2.1) & 75.8 (+10.0) & 68.8 (+6.4) & 87.4 (+4.0) \\
            %\noalign{\smallskip}
            %& $R_{P_\text{orb}}$\: [$\%$] & 2.8 & 1.9 & 3.3 & 3.1 & 2.1 & 5.8 \\
            \noalign{\smallskip}
             \cmidrule(lr){2-8}
            Active star & N$_{\rm FAP<1\%}$ & 48.0$\%$ & 95.2$\%$ & 34.8$\%$ & 88.4$\%$ & 98.0$\%$ & 65.5$\%$ \\
            \noalign{\smallskip}
            & N$_{\rm FAP\: 1-10\%}$ & 29.2$\%$ & 2.2$\%$ & 40.0$\%$ & 5.9$\%$ & 1.0$\%$ & 28.2$\%$ \\
            \noalign{\smallskip}
            & $C_{P_\text{rot}}$ ($C_{P_{\rm rot}/2}$)\tnote{3}\: [$\%$] & 4.8 (0.3) & 7.4 (0.8) & 11.7 (0.5) & 59 (5.6) & 48.5 (3.8) & 63.3 (0.8) \\
            \noalign{\smallskip}
            & $C_{P_\text{orb}}$\: [$\%$] & 0.3 & 0.6 & 0.6 & 0.1 & 0.4 & 0.3 \\
            \noalign{\smallskip}
            & $R$ ($+{P_{\rm rot}/2}$)\tnote{3}\: [$\%$] & 10.5 (+0.6) & 8.4 (+0.9) & 34.5 (+1.1) & 66.9 (+6.3) & 49.9 (+3.9) & 96.6 (+1.2) \\
            %\noalign{\smallskip}
            %& $R_{P_\text{orb}}$\: [$\%$] & 0.6 & 0.6 & 0.7 & 0.1 & 0.4 & 0.1 \\
            \noalign{\smallskip}
            \hline       
            \hline
     \end{tabular}  
     \begin{tablenotes}
     \item[1] The false alarm probability (FAP) has been evaluated through a bootstrap with replacement.
     \item[2] The false alarm probability (FAP) is the one provided by the code.
     \item[3] The percentage in parenthesis is related to the first harmonic of $P_{\rm rot}$. For FREDEC, this percentage corresponds to the number of datasets for which only $P_{\rm rot/2}$ is found as significant peak in the periodogram, while $P_{\rm rot}$ is not.
   \end{tablenotes}
 %    }
  \end{threeparttable}
\end{table*}

\section{Summary}
In this work we presented the results of a statistical analysis of an extended sample of simulated radial velocity datasets\footnote{The simulated datasets are freely available for testing other analysis tools and methods, upon request to the authors.} including stellar activity and planetary signals, devised as representations of realistic measurements for typical spectroscopic exoplanet surveys, in terms of state-of-the-art instrumentation and observing sampling. We simulated different levels and variability timescales of the stellar activity RV term, assumed as quasi-periodic correlated signal, with two goals: \textit{i)} testing the performances of widespread RV modelling techniques in retrieving Doppler signals of small semi-amplitude caused by low-mass planets (Sect. \ref{sect:results}), and \textit{ii)} exploring the properties of periodograms calculated with three different algorithms, in order to statistically characterize and compare their performances as a function of the stellar activity properties (Sect. \ref{sect:periodanalysis}). By pursuing the first goal, we assume that the existence of the planet is already demonstrated. This is the case of planets discovered with the transit technique, for which a mass estimate is pursued. 

As for the second goal, since we assume that nothing is known a-priori, both for the stellar activity and planetary signals, that is a problem of \textit{blind search}, that we have already dealt with in \citet{pinamontietal2017} for different scenarios. As done in that work, we tested the performances of the GLS, BGLS and FREDEC algorithms for a blind search of planetary signals in presence of stellar activity with well defined properties. Our mock datasets represent a hard challenge for these algorithms as the injected Doppler signal, which is consistent to what is expected for terrestrial planets, is small compared to the activity signal. Based on our results, there is no unique algorithm which out-performs the others in this case, and the discussion has been necessarily focused on how the stellar signals are recovered. However, within the same general framework we defined in this work, new simulations can be devised to move the focus on planetary signals, for instance by increasing their semi-amplitude, or considering more precise RVs, or simulating very-low activity stars.   

We analysed specific cases and presented the results schematically, in order to provide quick references to those interested in modelling RV time series for planetary studies. Tables \ref{Table:mcmcanalysis}, \ref{Table:periodanalysis1} and \ref{Table:periodanalysis2} contain the results for all the addressed scenarios. Among them, we highlight the following outcomes:
\begin{itemize}
    \item generally, the datasets containing more stable activity signals ($P_{\rm rot}\ll$ $\tau_{\rm AR}$) are easier to model, with improved performances in retrieving planetary signals;
    \item the correlation time scale $\tau_{\rm AR}$ is not always well recovered. When not much data are available (e.g. one semester of observations) it can be easily overestimated, especially in the case of stars with low and quickly variable activity. A tendency toward overestimation is also seen for active stars, but this is less pronounced. This result should be kept in mind when interpreting the fitted values for $\tau_{\rm AR}$ astrophysically, in particular when there are not ancillary data available for supporting the reality of long correlation time scale observed in the RVs;  
    \item the 68.3$^{th}$ percentile appears as a more accurate estimate for $K_{\rm b}$ than the 50$^{th}$ percentile for low-activity stars when $K_{\rm b,\:inj}$=1 $\ms$, $P_{\rm rot}\neq P_{\rm orb}$ and with two semesters of observations. Same happens for the selected scenarios with $K_{\rm b,\:inj}>$1 $\ms$ discussed in Sect. \ref{subsec:higherk};
    \item for stars with low activity levels and $P_{\rm rot}$ $\neq$ $P_{\rm orb}$, Bayesian model comparison shows that in general the use of the GP regression provides a significantly better fit to the data, especially with observations spanning two semesters, but this does not always correspond to a better identification of the planetary semi-amplitude, in the studied case of small-amplitude signals; %not correspond to better results in terms of planetary semi-amplitude for $K_{\rm b,\: inj}$=1 \ms;
    \item the different algorithms for periodogram analysis studied in Sect. \ref{sect:periodanalysis} follow similar behaviours depending on the hyper-parameters of the stellar activity model, being most effective when the star is quiet and $\tau_{\rm AR}$ is long;
    \item highly variable stellar activity (short $\tau_{\rm AR}$) affects significantly the performances of algorithms to search for periodic signals, especially for active stars. Their reliability, as defined in Table \ref{Table:analysedparamperiodogram}, is not very high due to the presence of a significant number of false positive signals, that does not decrease after increasing the number of measurements;
    \item in the cases with strong and highly variable stellar activity, which produces the most complex signal structures, the multi-frequency approach applied by FREDEC proves to be most effective, as proved by the higher $R$ value with respect to the other algorithms. This is consistent with the findings of \citet{pinamontietal2017}, which highlighted FREDEC to be less prone to false detections in the analysis of time series containing multiple complex signals.
\end{itemize}

It is worth noticing that the performances of all the periodogram algorithms in retrieving the planetary signals are quite poor in all the analysed scenarios, with $C_{Porb} \lesssim 5 \%$. This once again highlights the importance of applying high-level analysis techniques when trying to model planetary signals around active stars \citep[e.g. see][and references therein, for a review of different methods to deal with stellar activity]{dumusque17}.

Even though they must be considered mostly valid within the framework defined for our simulations, these results could be of some interest in a more general context, and could be used for devising diversified statistical studies aimed at optimizing follow-up strategies for different targets, number of planets and system architectures. They can be also extended to next generation high-resolution spectrographs, as for instance the operating VLT/ESPRESSO \citep{pepe14}, the upcoming EXPRES \citep{jurge16}, NEID \citep{schwab16}, and ELT-HIRES \citep{marconi16}, or NIR spectrographs such as CARMENES \citep{quirrenbach16}, SPIRou \citep{artigau14}, HPF \citep{mahadevan14}, or NIRPS \citep{wildi17}, that are expected to provide more precise RV measurements. 
Presently, the operating TESS mission is already detecting new transiting planets, and it is expected to discover $\sim$1\,000 planets with R<4 $\rearth$ at short-time cadence (2-minutes), 250 of which smaller than 2 $\rearth$ \citep{barclay18}. Taking into account planets that can actually be characterizable, i.e. are most suitable for mass measurements through precise radial velocity observations, \cite{barclay18} predict a distribution of detected orbital periods with a median of 7-8 days, likely less than 29-35 days at 95$\%$ significance level. Therefore, our simulations are particularly useful for planning the spectroscopic follow-up of small-size, low-mass planets discovered right now by TESS, and in the near future by the PLATO 2.0 mission \cite{rauer14}.  

\section*{Acknowledgements}
We thank the anonymous referee for helpful comments and suggestions. MD acknowledges financial support from Progetto Premiale 2015 FRONTIERA (OB.FU. 1.05.06.11) funding scheme of the Italian Ministry of Education, University, and Research. MP acknowledges support from the European Union Seventh Framework Programme (FP7/2007-2013) under grant agreement number 313014 (ETAEARTH). GS acknowledges financial support from "Accordo ASI-INAF" No. 2013-016-R.0 July 9, 2013 and July 9, 2015. We acknowledge the computing centres of INAF - Osservatorio Astronomico di Trieste / Osservatorio Astrofisico di Catania, under the coordination of the CHIPP project, for the availability of computing resources and support. We thank J\~{o}ao Faria (Instituto de Astrof\'{i}sica e Ci\^{e}ncias do Espa\c{c}o, Porto) for useful discussions.
%%%%%%%%%%%%%%%%%%%%%%%%%%%%%%%%%%%%%%%%%%%%%%%%%%



\bibliographystyle{mnras}
\bibliography{ref.bib} % if your bibtex file is called example.bib

%\begin{thebibliography}{99}
%\end{thebibliography}

%%%%%%%%%%%%%%%%%%%%%%%%%%%%%%%%%%%%%%%%%%%%%%%%%%

%%%%%%%%%%%%%%%%% APPENDICES %%%%%%%%%%%%%%%%%%%%%

\appendix

\section{Simulations properties}

Table \ref{Table:simuldistr} shows some relations existing between the injected model parameters considered in this work, generated as described in Sect. \ref{sect:simsetup}.

\begin{table*}
\caption[]{Relations between some of the injected model parameters considered in this work, and typical RMS of the datasets. Each value is given as $16^{\rm th}$, $50^{\rm th}$ and $84^{\rm th}$ percentiles of each distribution consisting of 5\,000 samples.} 
\begin{threeparttable}
         \label{Table:simuldistr}
   \normalsize
%   \makebox[\textwidth][c]{
   \begin{tabular}{cccccc}
            \hline
            \hline
            \noalign{\smallskip}
             & & \multicolumn{2}{c}{\textbf{P$_{\rm rot}\sim$P$_{\rm orb}$}} & \multicolumn{2}{c}{\textbf{P$_{\rm rot}$ $\neq$ P$_{\rm orb}$} }\\
            \cmidrule(lr){3-4}\cmidrule(lr){5-6}
            \noalign{\smallskip}
           % \hline
            & Parameter & short $\tau_{\rm AR}$ & long $\tau_{\rm AR}$ & short $\tau_{\rm AR}$ & long $\tau_{\rm AR}$ \\
         %   \hline
        %  \cmidrule(lr){3-4}\cmidrule(lr){5-6}
          \noalign{\smallskip}
            \noalign{\smallskip}
            \hline \hline
            \noalign{\smallskip}
            \multicolumn{6}{l}{\textbf{One observing season}} \\ % & & & & & & & & & & & & & \\
            \noalign{\smallskip}
            \hline
            \noalign{\smallskip}
            Low activity star & $P_{\rm rot}$-$P_{\rm orb}$ [d] & -0.03$\pm$2.13 & -0.1$\pm$2.1 & 6.9$\pm$2.1 & 7.1$\pm$2.1 \\
            \noalign{\smallskip}
            & $\tau_{\rm AR}$-$P_{\rm rot}$ [d] & 0.0016$\pm$2.1 & 179.9$^{\rm +10.4}_{\rm -9.6}$ & -0.006$\pm$2.1 & 179.7$^{\rm +10.4}_{\rm -10.2}$ \\
            %\noalign{\smallskip}
            %& $\tau_{\rm AR}$-$P_{\rm orb}$ [d] & -0.0083$\pm$2.1 & 179.9$^{\rm +10.2}_{\rm -9.8}$ & 6.9$\pm$2.1 & 186.8$^{\rm +10.4}_{\rm -10.2}$ \\
            \noalign{\smallskip}
            & $\text{RV}_\text{rms}$ [\ms] & 3.4$^{\rm +0.7}_{\rm -0.6}$ & 2.9$^{\rm +0.8}_{\rm -0.5}$ & 3.4$^{\rm +0.7}_{\rm -0.5}$ & 2.9$^{\rm +0.8}_{\rm -0.5}$ \\
            \noalign{\smallskip}
          %   \cmidrule(lr){2-14}
            Active star & $P_{\rm rot}$-$P_{\rm orb}$ [d] &  0.02$\pm$1.4 & -0.03$^{\rm +1.43}_{\rm -1.39}$ & -8.0$\pm$1.8 & -8.0$\pm$1.8 \\
            \noalign{\smallskip}
            & $\tau_{\rm AR}$-$P_{\rm rot}$ [d] & -0.005$\pm$1.4 & 90.4$^{\rm +10.1}_{\rm -10.2}$ & 0.02$\pm$1.4 & 90.2$^{\rm +9.7}_{\rm -9.9}$  \\
            %\noalign{\smallskip}
            %& $\tau_{\rm AR}$-$P_{\rm orb}$ [d] & -0.03$^{\rm +1.45}_{\rm -1.38}$ & 90.2$\pm$10.2 & -8.0$\pm$1.8 & 82.1$^{\rm +9.7}_{\rm -9.9}$ \\
            \noalign{\smallskip}
            & $\text{RV}_\text{rms}$ [\ms] & 14.2$^{\rm +2.6}_{\rm -2.3}$ & 10.9$^{\rm +4.1}_{\rm -3.3}$ & 14.1$^{\rm +2.6}_{\rm -2.3}$ & 10.9$^{\rm +3.9}_{\rm -3.2}$ \\
            \noalign{\smallskip}
            \hline
            \noalign{\smallskip}
            \multicolumn{6}{l}{\textbf{Two observing seasons}} \\
            \noalign{\smallskip}
            \hline
            \noalign{\smallskip}
            Low activity star & $P_{\rm rot}$-$P_{\rm orb}$ [d] & -0.04$\pm$2.2 & 0.03$\pm$2.1 & 7.0$\pm$2.1 & 6.9$\pm$2.1 \\
            \noalign{\smallskip}
            & $\tau_{\rm AR}$-$P_{\rm rot}$ [d] & 0.06$\pm$2.1 & 180.1$^{\rm +10.4}_{\rm -10.1}$ & 0.008$\pm$2.1 & 180.3$\pm$10.1 \\
            %\noalign{\smallskip}
            %& $\tau_{\rm AR}$-$P_{\rm orb}$ [d] & -0.009$\pm$2.1 & 180.2$^{\rm +10.2}_{\rm -10.1}$ & 6.9$\pm$2.2 & 187.2$^{\rm +10.1}_{\rm -10.2}$ \\
            \noalign{\smallskip}
            & $\text{RV}_\text{rms}$ [\ms] & 3.5$^{\rm +0.6}_{\rm -0.5}$ & 3.2$^{\rm +0.7}_{\rm -0.6}$ & 3.5$^{\rm +0.6}_{\rm -0.5}$ & 3.2$^{\rm +0.7}_{\rm -0.5}$ \\
            \noalign{\smallskip}
          %   \cmidrule(lr){2-14}
            Active star & $P_{\rm rot}$-$P_{\rm orb}$ [d] & -0.01$\pm$1.4 & 0.04$^{\rm +1.39}_{\rm -1.42}$ & -8.0$\pm$1.8 & -8.0$^{\rm +1.8}_{\rm -1.7}$ \\
            \noalign{\smallskip}
            & $\tau_{\rm AR}$-$P_{\rm rot}$ [d] & 0.0035$\pm$1.4 & 90.2$^{\rm +9.9}_{\rm -10.0}$ & 0.0006$\pm$1.4 & 89.9$\pm$9.7 \\
            %\noalign{\smallskip}
            %& $\tau_{\rm AR}$-$P_{\rm orb}$ [d] & 0.007$\pm$1.4 & 90.2$^{\rm +9.8}_{\rm -10.1}$ & -8.0$^{\rm +1.9}_{\rm -1.8}$ & 81.9$^{\rm +10.1}_{\rm -9.8}$ \\
            \noalign{\smallskip}
            & $\text{RV}_\text{rms}$ [\ms] & 14.7$^{\rm +2.0}_{\rm -1.8}$ & 13.0$^{\rm +3.4}_{\rm -3.0}$ & 14.7$^{\rm +1.9}_{\rm -1.8}$ & 12.9$^{\rm +3.3}_{\rm -2.9}$ \\
            \noalign{\smallskip}
            \hline       
            \hline
     \end{tabular}  
  \end{threeparttable}
\end{table*}

\section{Assessing possible biases introduced by the observing calendar}
\label{appx2}
\begin{figure}
  \label{medianwf} 
  \includegraphics[width=\hsize]{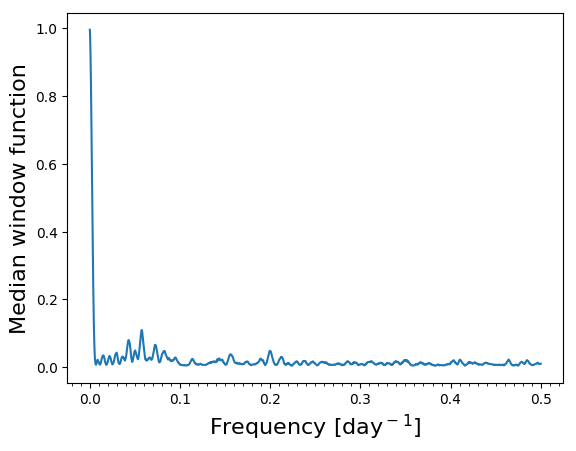}
  \includegraphics[width=\hsize]{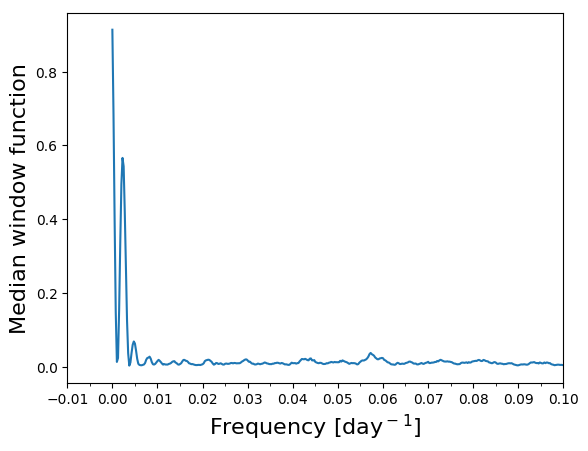} 
  \caption{Examples of median window functions for different simulated scenarios. \textit{Upper panel}: quiet star with $P_{\rm rot}\sim P_{\rm orb}$, long $\tau_{\rm AR}$, and one season of observations. \textit{Lower panel}: active star with $P_{\rm rot}\neq P_{\rm orb}$, long $\tau_{\rm AR}$, and two seasons of observations.}
\end{figure}

\begin{figure}
  \label{meanphasecoverage} 
  \includegraphics[width=\hsize]{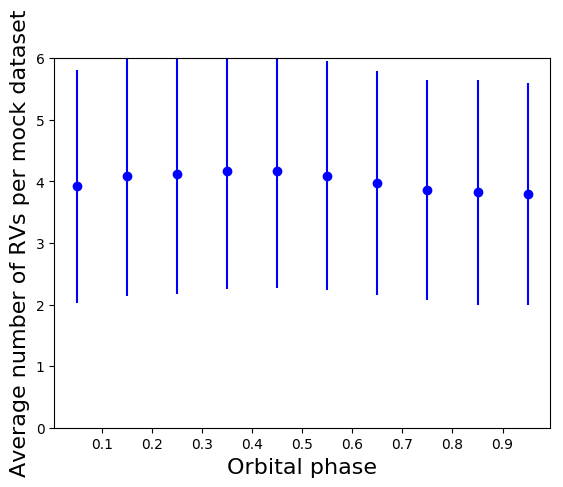}
  \includegraphics[width=\hsize]{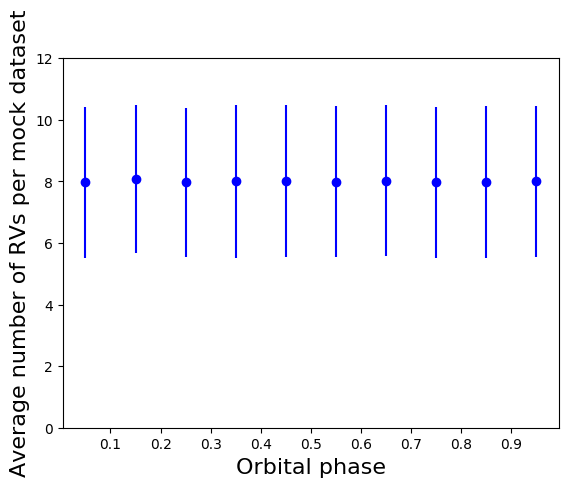} 
  \caption{Examples showing how the number of RV data distributes on average (per single RV dataset) within phase bins of size 0.1, for different simulated scenarios. \textit{Upper panel}: quiet star with $P_{\rm rot}\sim P_{\rm orb}$, long $\tau_{\rm AR}$, and one season of observations. \textit{Lower panel}: quiet star with $P_{\rm rot}\neq P_{\rm orb}$, long $\tau_{\rm AR}$, and two seasons of observations. The error bars represent the RMS of the number of RV measurements within each phase bin.}
\end{figure}

We verified whether our particular choice of the 63-nights calendar could have introduced biases in our results through three different checks: \textit{i}) we calculated the median window functions for each scenario, and they do not show significant peaks at specific frequencies (except for the expected 1-year alias for $N_{\rm epochs, s2}$ data), which would indicate the presence of aliases introduced by the sampling (Fig. \ref{medianwf}); \textit{ii}) we found that the orbits of the simulated planets are on average uniformly covered in phase by the RV measurements, i.e. there are no parts of the orbit systematically over/under sampled (Fig. \ref{meanphasecoverage}); \textit{iii}) we simulated 200 RV datasets by randomly generated 40 epochs from a uniform distribution over a range of one semester, and then adding 40 more epochs randomly drawn within a time range shifted by 365 days. This implies that each dataset has a different observing calendar which does not stem from the same array of epochs. We considered the case of a quiet star with $P_{\rm rot}\neq P_{\rm orb}$, long $\tau_{\rm AR}$, and two seasons of observations. We found that the results of a GP-based MC analysis are similar to those presented in Sect. \ref{sec:mcanalysis}, indicating that the latter are free from biases due to a particular choice of the mold calendar.

%%%%%%%%%%%%%%%%%%%%%%%%%%%%%%%%%%%%%%%%%%%%%%%%%%

% Don't change these lines
\bsp	% typesetting comment
\label{lastpage}
\end{document}